\documentstyle[preprint,tighten,aps,colordvi,epsfig,psfig]{revtex}
\def\slashed{{/}\mskip-10.0mu}

\begin{document}
\newcommand{\real}{{\sf I}\kern-.12em{\sf R}}
\newcommand{\fillrhd}{{\rhd\kern-.81em\bullet}\kern-.47em\triangleright}
\draft
\vskip 2cm

\title{Two-loop renormalization of scalar and pseudoscalar fermion bilinears 
on the lattice}

\author{A. Skouroupathis and H. Panagopoulos}
\address{Department of Physics, University of Cyprus, P.O. Box 20537,
Nicosia CY-1678, Cyprus \\
{\it email: }{\tt php4as01@ucy.ac.cy, haris@ucy.ac.cy}}
\vskip 3mm

\date{\today}

\maketitle

\begin{abstract}

We compute the two-loop renormalization functions, in the RI\,$^\prime$ 
scheme, of local bilinear quark operators $\bar{\psi}\Gamma\psi$, where 
$\Gamma$ denotes the Scalar and Pseudoscalar Dirac matrices, in the 
lattice formulation of QCD. We consider both the flavor non-singlet and 
singlet operators; the latter, in the scalar case, leads directly to the
two-loop fermion mass renormalization, $Z_m$. 

As a prerequisite for the above, we also compute the quark field 
renormalization, $Z_{\psi}$, up to two loops.

We use the clover action for fermions and the Wilson action for gluons. 
Our results are given as a polynomial in $c_{SW}$, in terms of both the
renormalized and bare coupling constant, in the renormalized Feynman gauge. 
We also confirm the 1-loop renormalization functions, for generic gauge.

Finally, we present our results in the $\overline{MS}$ scheme, for easier
comparison with calculations in the continuum.

The corresponding results, for fermions in an {\it arbitrary}
representation, are included in an Appendix.


\end{abstract}

\newpage


\section{Introduction}
\label{introduction}

Studies of hadronic properties using the lattice formulation of QCD rely on
the computation of matrix elements and correlation functions of composite 
operators, made out of quark fields. A whole variety of such operators has
been considered and studied in numerical simulations, including local and
extended bilinears, and four-fermi operators. 
A proper renormalization of these operators is most often indispensable for 
the extraction of physical results from the lattice.

In this work we study the renormalization of fermion bilinears 
${\cal O}=\bar{\psi}\Gamma\psi$ on the lattice, where $\Gamma =
\openone,\,\gamma_5$. 
We consider both flavor singlet and nonsinglet operators.
The cases $\Gamma=\gamma_{\mu},\,\gamma_5\,\gamma_{\mu},\,\gamma_5\,\sigma_{\mu\,\nu}$, 
will be presented in a sequel to this work.
In order to obtain the  renormalization functions of 
fermion bilinears we also compute the quark 
field renormalization, $Z_{\psi}$, as a prerequisite. 

We employ the standard Wilson action for gluons and clover-improved
Wilson fermions. The number of quark flavors $N_f$, the number 
of colors $N_c$ and the clover coefficient $c_{{\rm SW}}$ are kept as 
free parameters.

Our two-loop calculations have been performed in the bare and in 
the renormalized Feynman gauge. For 1-loop quantities, the 
gauge parameter is allowed to take arbitrary values.

The main results presented in this work are the following 
2-loop bare Green's functions (amputated, one-particle irreducible (1PI)): 
\begin{itemize}
\item Fermion self-energy: $\Sigma^L_{\psi}(q,a_{_{\rm L}})$ 
\item 2-pt function of the scalar $\bar{\psi}\psi:
$ $\Sigma^L_S(q a_{_{\rm L}})$ 
\item 2-pt function of the pseudoscalar $\bar{\psi}\gamma_5\psi:
$ $\Sigma^L_P(q a_{_{\rm L}})$  
\end{itemize}
($a_{_{\rm L}}\,:$ lattice spacing, $q:$ external momentum)

In general, one can use bare Green's functions to construct 
$Z_{{\cal O}}^{X,Y}$, the renormalization function for 
operator ${\cal O}$, computed within a regularization $X$ 
and renormalized in a scheme $Y$. 

We employ two widely used schemes to compute the various 
2-loop renormalization functions: 
\begin{itemize}
\item The $RI^{\prime}$ scheme: $Z_{\psi}^{L,RI^{\prime}}$, 
$Z_S^{L,RI^{\prime}}$, $Z_P^{L,RI^{\prime}}$ 
\item The $\overline{MS}$ scheme: $Z_{\psi}^{L,\overline{MS}}$, 
$Z_S^{L,\overline{MS}}$, $Z_P^{L,\overline{MS}}$
\end{itemize}

The flavor singlet scalar renormalization function is equal 
to the fermion mass multiplicative renormalization, $Z_m$, 
which is an essential ingredient in computing quark masses.

For convenience, the results for $Z_{{\cal O}}^{X,Y}$
are given in terms of both the bare coupling
constant $g_{\rm o}$ and the renormalized one: $g_{RI^{\prime}}$\,, 
$g_{\overline{MS}}$. 

Finally, as one of several checks on our results, we construct
the 2-loop renormalized Green's functions in  $RI^{\prime}$: 
$\Sigma_{{\cal O}}^{RI^{\prime}}(q,\bar{\mu})$ (${\cal O}\equiv\psi,S,P$),
as well as their counterparts in $\overline{MS}$:
$\Sigma_{{\cal O}}^{\overline{MS}}(q,\bar{\mu})$.
The values of all these functions, computed on the lattice, 
coincide with values computed in dimensional regularization
(as can be inferred, e.g., from ~\cite{Gracey}).

The present work is the first two-loop computation of the 
renormalization of fermion bilinears on the lattice. 
One-loop computations of the same quantities exist for quite some time
now (see, e.g., \cite{MZ}, \cite{Aoki98}, \cite{Capitani} and references therein).
There have been made several attempts to estimate 
$Z_{{\cal O}}$ non-perturbatively; recent
results can be found in Refs. 
\cite{Zhang,Becirevic,Aoki,Galletly,Sommer,DellaMorte}.
Some results have also been obtained using stochastic perturbation
theory~\cite{DiRenzo}. A related computation, regarding the fermion mass
renormalization $Z_m$ with staggered fermions can be found in~\cite{Trottier}. 

The paper is organized as follows: Section \ref{Formulation} provides a
formulation of the problem, along with all necessary definitions of
renormalization schemes and of the quantities to compute. Section 
\ref{Results} describes our computational methods and the results
which are obtained. Finally, in Section \ref{Discussion} we discuss
some salient features of our calculation, and comment on future
extensions to the present work.

Recently, there has been some interest in gauge theories with
fermions in representations other than the fundamental. Such theories
are being studied in various contexts \cite{KUY,GG,CDDLP,BPV,GK,EHS}, e.g.,
supersymmetry, phase transitions, and the `AdS/QCD' correspondence. It is
relatively straightforward to generalize our results to an arbitrary
representation; this is presented in the Appendix.


\section{Formulation of the problem}
\label{Formulation}

\subsection{Lattice action} 

We will make use of the Wilson formulation of the QCD action on the
lattice, with the addition of the clover (SW)~\cite{SW}
term for fermions. In standard notation, it reads:

\begin{eqnarray}
S_L &=& S_G + \sum_{f}\sum_{x} (4r+m_{\rm o})\bar{\psi}_{f}(x)\psi_f(x)
\nonumber \\
&-& {1\over 2}\sum_{f}\sum_{x,\,\mu}
\bigg{[}\bar{\psi}_{f}(x) \left( r - \gamma_\mu\right)
U_{x,\,x+\mu}\,\psi_f(x+{\mu}) \nonumber \\
&~& \hspace{1.7cm}+\bar{\psi}_f(x+{\mu})\left( r + \gamma_\mu\right)
U_{x+\mu,\,x}\,\psi_{f}(x)\bigg{]}\nonumber \\
&+& {i\over 4}\,c_{\rm SW}\,\sum_{f}\sum_{x,\,\mu,\,\nu} \bar{\psi}_{f}(x)
\,\sigma_{\mu\nu} \,{\hat F}_{\mu\nu}(x) \,\psi_f(x),
\label{latact}
\end{eqnarray}
\begin{eqnarray}
{\rm where:}\qquad {\hat F}_{\mu\nu} &\equiv& {1\over{8a^2}}\,
(Q_{\mu\nu} - Q_{\nu\mu})\\
{\rm and:\qquad} Q_{\mu\nu} &=& U_{x,\, x+\mu}\,U_{x+\mu,\, x+\mu+\nu}\,U_{x+\mu+\nu,\, x+\nu}
\,U_{x+\nu,\, x}\nonumber \\
&+& U_{ x,\, x+ \nu}\,U_{ x+ \nu,\, x+ \nu- \mu}\,U_{ x+ \nu- \mu,\, x- \mu}\,U_{ x- \mu,\, x} \nonumber \\
&+& U_{ x,\, x- \mu}\,U_{ x- \mu,\, x- \mu- \nu}\,U_{ x- \mu- \nu,\, x- \nu}\,U_{ x- \nu,\, x}\nonumber \\
&+& U_{ x,\, x- \nu}\,U_{ x- \nu,\, x- \nu+ \mu}\,U_{ x- \nu+ \mu,\, x+ \mu}\,U_{ x+ \mu,\, x}
\label{latact2}
\end{eqnarray}

$S_G$ is the standard pure gluon action, made out of $1{\times}1$
plaquettes. The clover coefficient $c_{\rm SW}$ is treated here as a free parameter;
$r$ is the Wilson parameter (set to $r=1$ henceforth);
$f$ is a flavor index; $\sigma_{\mu\nu} =(i/2) [\gamma_\mu,\,\gamma_\nu]$.
Powers of the lattice spacing $a_{_{\rm L}}$ have been omitted and may be
directly reinserted by dimensional counting. 

The ``Lagrangian mass'' $m_{\rm o}$ is a free parameter here. However, since we 
will be using mass independent renormalization schemes, all renormalization functions
which we will be calculating, must be evaluated at vanishing renormalized mass, that
is, when $m_{\rm o}$ is set equal to the critical value 
$m_{\rm cr}$: $m_{\rm o}\to m_{\rm cr}=0+{\cal O}(g_{\rm o}^2)$.

\subsection{Definition of renormalized quantities}
 
As a prerequisite to our programme, we will need the renormalization
functions for the gluon, ghost and fermion fields ($A_\mu^a,\ c^a, \ \psi$), and for the
coupling constant $g$ and gauge parameter $\alpha$, defined as follows:
\begin{eqnarray}
A_{\mu\,{\rm o}}^a &=& \sqrt{Z_A}\,A^a_{\mu}, \hspace{1cm} c^a_{\rm o}=\sqrt{Z_c}\,c^a, \hspace{1cm} 
\psi_{\rm o}=\sqrt{Z_{\psi}}\,\psi \nonumber \\
&&\hskip 1cm g_{\rm o} = \mu^{\epsilon}\,Z_g\,g, \hspace{1cm} \alpha_{\rm o} = Z_a^{-1}\,Z_A\,\alpha
\label{fields}     
\end{eqnarray}

The value of each $Z_{{\cal O}}$ depends both on the regularization $X$ and on 
the renormalization scheme $Y$ employed, and thus should properly be denoted as
$Z^{X,Y}_{{\cal O}}$. The scale $\mu$ enters the relation between $g_o$ and $g$
only in dimensional regularization ($D=4-2\epsilon$ dimensions).

We will need $Z_A,\,Z_c,\,Z_{\alpha}\,{\rm and}\,Z_g$ to 1 loop and $Z_{\psi}$ to 
2 loops. Our 1-loop results, performed in a generic gauge, are in
agreement with results found in the literature (see, e.g., Refs.~\cite{Capitani,Bode}).

\subsection{Definition of the $\mathbf{RI^{\prime}}$ scheme} 

This renormalization scheme~\cite{Martinelli,Franco,Chetyrkin2000} is
more immediate for a lattice regularized theory. 
It is defined by imposing a set of normalization conditions on matrix elements
at a scale $\bar{\mu}$, where (just as in the $\overline{MS}$ scheme)~\cite{Collins}:

\begin{equation}
\bar{\mu}=\mu\,(4\pi/e^{\gamma_{\rm E}})^{1/2} 
\label{mubar}
\end{equation}
($\gamma_{{\rm E}}$ is the Euler constant).

In Euclidean space, the fermion self energy
$\Sigma^L_{\psi}(q,a_{_{\rm L}})=i\slashed{q}+m_{\rm o}+{\cal O}(g_o^2)$
is renormalized through:
\begin{equation}
\lim_{a_{_{\rm L}}\rightarrow 0} \left[Z_{\psi}^{L,RI^{\prime}}(a_{_{\rm L}}\bar{\mu})\,
{\rm tr}\left(\Sigma_{\psi}^L(q,a_{_{\rm L}})\,\slashed{q}\right)
/(4i\,q^2)\right]_{q^2=\bar{\mu}^2}=1
\label{ZpsiRule}
\end{equation}
The trace here is over Dirac indices; a Kronecker delta in color and
in flavor indices has been factored out of the definition of
$\Sigma_{\psi}^L$.

Similarly, for the ghost self energy $\Sigma^L_c(q,a_{_{\rm L}})=q^2+{\cal O}(g_o^2)$: 
\begin{equation}
\lim_{a_{_{\rm L}}\rightarrow 0} \left
    [Z_c^{L,RI^{\prime}}(a_{_{\rm L}}\bar{\mu})\frac{\Sigma^L_c(q,a_{_{\rm L}})}{q^2}\right ]_{q^2=\bar{\mu}^2}=1
\label{ZcRule}
\end{equation}

$Z_A$ and $Z_\alpha$ are extracted from the gluon propagator $G^L_{\mu\,\nu}(q,a_{_{\rm L}})$ 
with radiative corrections\footnote{One should carefully distinguish among the 
following standard symbols:
$a_{_{\rm L}}$: lattice spacing;  $\alpha_{\rm o}$, $\alpha_{RI^{\prime}}$, 
$\alpha_{\overline{MS}}$: bare and renormalized gauge parameters.}:

\begin{equation}
G^L_{\mu\,\nu}(q,a_{_{\rm L}})=\frac{1}{q^2}\left [\frac{\delta_{\mu\,\nu}-q_{\mu}q_{\nu}/q^2}{\Pi_T(a_{_{\rm L}}q)}
+\alpha_{\rm o}\,\frac{q_{\mu}q_{\nu}/q^2}{\Pi_L(a_{_{\rm L}}q)}\right ]
\label{GluonProp}
\end{equation}
where $\Pi_{T,L}(a_{_{\rm L}}q)=1+{\cal O}(g_o^2)$. The normalization conditions are: 

\begin{eqnarray}
\lim_{a_{_{\rm L}}\rightarrow 0} \left [ Z_A^{L,RI^{\prime}}(a_{_{\rm L}}\bar{\mu})\,\frac{1}{\Pi_T(a_{_{\rm L}}q)}\right ]_ 
{q^2=\bar{\mu}^2} &=& 1 \\
\lim_{a_{_{\rm L}}\rightarrow 0} \left [ Z_{\alpha}^{L,RI^{\prime}}(a_{_{\rm L}}\bar{\mu})\,\frac{1}{\Pi_L(a_{_{\rm L}}q)}\right ]_
{q^2=\bar{\mu}^2} &=& 1 
\label{ZAZaRules}
\end{eqnarray} 
We have checked explicitly that $Z^{L,RI^{\prime}}_{\alpha}=1$ 
up to one loop, in agreement with the continuum.

For consistency with the Slavnov-Taylor identities, 
$Z_g$ in the $RI^{\prime}$ scheme is defined as in the
$\overline{MS}$ scheme. In dimensional regularization ($DR$)
this is achieved by requiring that the gluon-fermion-antifermion
1PI vertex function, $G_{A\bar{\psi}\psi}$, renormalizes as 
follows~\cite{Gracey}: 
\begin{equation}
\lim_{\epsilon\rightarrow 0} \left[Z_{\psi}^{DR,RI^{\prime}}\,(Z_A^{DR,RI^{\prime}})^{1/2}
Z_g^{DR,RI^{\prime}}G_{A\bar{\psi}\psi}(q)\right]
_{q^2=\bar{\mu}^2} = G^{{\rm finite}}_{A\bar{\psi}\psi} 
\label{ZgRulesA} 
\end{equation}
The value of $Z_g$ is tuned in such a way as to absorb only the poles
in $\epsilon$ which appear in $G_{A\bar{\psi}\psi}$ (together with
matching powers of $\ln(4\pi)-\gamma_{\rm E}$); this leads to a
result for $G_{A\bar{\psi}\psi}^{{\rm finite}}$ which is finite
but not unity. Before rescaling, we have first divided 
$G_{A\bar{\psi}\psi}$ by the bare coupling constant, 
as in Ref.~\cite{LarinVermaseren}, in order to have
unity as the tree level value for 
$G_{A\bar{\psi}\psi}^{{\rm finite}}$. We have set the fermion momentum 
to zero; $q$ refers to the gluon/antifermion momentum. 
Alternatively, a similar procedure can be performed on the gluon-ghost-antighost vertex:
\begin{equation}
\lim_{\epsilon\rightarrow 0} \left[Z_c^{DR,RI^{\prime}}\,(Z_A^{DR,RI^{\prime}})^{1/2}
Z_g^{DR,RI^{\prime}}G_{A\bar{c}c}(q)\right]_{q^2=\bar{\mu}^2} = G^{{\rm finite}}_{A\bar{c}c}
\label{ZgRulesB}
\end{equation} 
Eq.(\ref{ZgRulesB}) leads to exactly the same value for $Z_g$.

The corresponding renormalization conditions on the lattice read:
\begin{equation}
\lim_{a_{_{\rm L}}\rightarrow 0} \left [Z_{\psi}^{L,RI^{\prime}}\,(Z_A^{L,RI^{\prime}})^{1/2}
Z_g^{L,RI^{\prime}}G^{L}_{A\bar{\psi}\psi}(q,a_{_{\rm L}})\right]
_{q^2=\bar{\mu}^2} = G^{{\rm finite}}_{A\bar{\psi}\psi}
\label{ZgRules2a}
\end{equation} 
or, equivalently: 
\begin{equation}
\lim_{a_{_{\rm L}}\rightarrow 0} \left[Z_c^{L,RI^{\prime}}\,(Z_A^{L,RI^{\prime}})^{1/2}
Z_g^{L,RI^{\prime}}G^{L}_{A\bar{c}c}(q,a_{_{\rm L}})\right]
_{q^2=\bar{\mu}^2} = G^{{\rm finite}}_{A\bar{c}c}
\label{ZgRules2b}
\end{equation} 
where the expressions $G^{{\rm finite}}_{A\bar{\psi}\psi}$ and
$G^{{\rm finite}}_{A\bar{c}c}$ are required to be the {\it same} as those
stemming from the continuum (Eqs.(\ref{ZgRulesA}),(\ref{ZgRulesB})). We have calculated
$Z_g^{L,RI^{\prime}}$, using either one of Eqs.(\ref{ZgRules2a}),
(\ref{ZgRules2b}), and have verified that the same result is obtained. 

\subsection{Conversion to the $\mathbf{\overline{MS}}$ scheme}
 
For easier comparison with calculations coming from the continuum, we need
to express our results in the $\overline{MS}$ scheme. Each
renormalization function on the lattice, $Z^{L,RI^{\prime}}_{{\cal O}}$, 
may be expressed as a power series in the renormalized
coupling constant $g_{RI^{\prime}}$. 
For the purposes of our work the conversion of $g_{RI^{\prime}}$ to 
$\overline{MS}$ is trivial since:

\begin{equation}
g_{RI^{\prime}}=g_{\overline{MS}}+{\cal O}(g^9_{\overline{MS}}) 
\label{aConversion}
\end{equation}

As already mentioned, our 1-loop calculations for $Z_A,\,Z_c,\,Z_{\alpha}$
and $Z_g$ are performed in a generic gauge, $\alpha_{RI^{\prime}}$.
The conversion to the $\overline{MS}$ scheme is given by~\cite{Retey}:

\begin{equation}
\alpha_{RI^{\prime}}=\frac{Z_A^{L,\overline{MS}}}{Z_A^{L,RI^{\prime}}}\,\alpha_{\overline{MS}}
\equiv \alpha_{\overline{MS}}\, /\, C_A(g_{\overline{MS}}, \alpha_{\overline{MS}})
\label{alphaConversion}
\end{equation}
Since the ratio of $Z$'s appearing in Eq.(\ref{alphaConversion}) must be
{\em regularization independent}, it may be calculated more easily in
dimensional regularization~\cite{Gracey}; to 1 loop, the conversion
factor  $C_A$ equals: 
\begin{equation}
C_A(g,\alpha)=\frac{Z_A^{DR,RI^{\prime}}}{Z_A^{DR,\overline{MS}}}=1+\frac{g^2}{36(16\pi^2)}\,\left[\left(9\alpha^2+ 
18\alpha +97\right)\,N_c - 40N_f\right]
\label{CA}
\end{equation}
(Here, and throughout the rest of this work, both $g$ and $\alpha$ are
in the $\overline{MS}$ scheme, unless specified otherwise.) 

Once we have computed the renormalization functions in the $RI^{\prime}$
scheme we can construct their $\overline{MS}$ counterparts using 
conversion factors which, up to the required perturbative order, are
given by:

\begin{eqnarray}
C_c(g,\alpha)&\equiv&\frac{Z_c^{L,RI^{\prime}}}{Z_c^{L,\overline{MS}}}=\frac{Z_c^{DR,RI^{\prime}}}{Z_c^{DR,\overline{MS}}}=1+\frac{g^2}{16\pi^2}\,N_c \label{Cc} \\
C_{\psi}(g,\alpha)&\equiv&\frac{Z_{\psi}^{L,RI^{\prime}}}{Z_{\psi}^{L,\overline{MS}}}=\frac{Z_{\psi}^{DR,RI^{\prime}}}{Z_{\psi}^{DR,\overline{MS}}} \nonumber \\
&=& 1 - \frac{g^2}{16\pi^2}\,\,c_F\,\alpha 
+\frac{g^4}{8\,(16\pi^2)^2}\,c_F\bigg[\left(8\alpha^2 + 5\right)\,c_F +  14\,N_f \nonumber \\ 
&& \hskip 3.5cm -\left(9\alpha^2-24\zeta(3)\,\alpha+52\alpha-24\zeta(3)+82\right)\,N_c\bigg] 
\label{Cpsi}
\end{eqnarray}
where $c_F= (N_c^2-1)/(2\,N_c)$ is the quadratic Casimir operator in the fundamental 
representation of the color group; $\zeta(x)$ is Riemann's zeta
function. (We employ a standard normalization for the generators of
the algebra, $T^a$ \,, see the Appendix.)

\subsection{Renormalization of fermion bilinears} 

The lattice operators ${\cal O}_{\Gamma}=\bar{\psi}\,\Gamma\,\psi$ must, in general, 
be renormalized in order to have finite matrix elements. 
We define
renormalized operators by 
\begin{equation}
{\cal
  O}^{RI^{\prime}}_{\Gamma}=Z^{L,RI^{\prime}}_{\Gamma}(a_{_{\rm L}}\bar\mu)\,{\cal
  O}_{\Gamma\,{\rm o}}
\label{RenormOper}
\end{equation}

The flavor singlet scalar operator receives also an additive
renormalization, which must be taken into account; we discuss this
issue in the following subsection.
For the scalar (S) and pseudoscalar (P) operators, the renormalization
functions $Z_{\Gamma}^{L,RI^{\prime}}$ can be obtained  
through the corresponding bare 2-point functions $\Sigma^L_{\Gamma}(q a_{_{\rm L}})$
(amputated, 1PI) on the lattice, in the following way: 

\begin{eqnarray}
\lim_{a_{_{\rm L}}\rightarrow 0}\left[Z_{\psi}^{L,RI^{\prime}}\,Z_S^{L,RI^{\prime}}\,\Sigma^L_S(q a_{_{\rm L}})\right]_{q^2=\bar{\mu}^2} &=& \openone \label{ZXrulesa} \\
\lim_{a_{_{\rm L}}\rightarrow 0}\left[Z_{\psi}^{L,RI^{\prime}}\,Z_P^{L,RI^{\prime}}\,\Sigma^L_P(q a_{_{\rm L}})\right]_{q^2=\bar{\mu}^2} &=& \gamma_5
\label{ZXrulesb}
\end{eqnarray}
where:
\begin{equation}
\Sigma^L_S(q a_{_{\rm L}})=\openone+{\cal O}(g_{\rm
  o}^2)\quad,\quad\Sigma^L_P(q a_{_{\rm L}})=\gamma_5+{\cal O}(g_{\rm o}^2)
\label{2ptFunct}
\end{equation}

Once the quantities $Z_\Gamma^{L,RI^{\prime}}$ have been calculated, one may
proceed to compute them also in the $\overline{MS}$ scheme. In the
case of the scalar operator (${\cal O}_{S\,{\rm o}}=\bar{\psi}_{\rm o}\psi_{\rm o}$), the  
renormalization function, $Z_S^{L,\overline{MS}}$, can be obtained by:
\begin{equation}
Z_S^{L,\overline{MS}}=Z_S^{L,RI^{\prime}} / C_S(g,\alpha)
\label{ScalarConversion}
\end{equation}
where $C_S(g,\alpha)$ is a {\em regularization independent} conversion
factor and has been calculated in dimensional regularization~\cite{Gracey}:
\begin{eqnarray}
C_S(g,\alpha)&\equiv&\frac{Z_S^{L,RI^{\prime}}}{Z_S^{L,\overline{MS}}}=
\frac{Z_S^{DR,RI^{\prime}}}{Z_S^{DR,\overline{MS}}} \nonumber \\
&=& 1+\frac{g^2}{16\pi^2}\,c_F\,\left(\alpha + 4\right) 
+\frac{g^4}{24\,(16\pi^2)^2}\,c_F\,\bigg[\left(24\alpha^2+96\alpha-288\zeta(3) +  57\right)\,c_F + 166\,N_f \nonumber \\
&& \hskip 4.5cm - \left(18\alpha^2 + 84\alpha -  432\zeta(3) + 1285\right)\,N_c\bigg] 
\label{CS}
\end{eqnarray}

The treatment of the pseudoscalar operator (${\cal
O}_{P\,{\rm o}}=\bar{\psi}_{\rm o}\gamma_5\psi_{\rm o}$) in the $\overline{MS}$ scheme
requires special attention, due to the non-unique generalization of
$\gamma_5$ to D dimensions. A practical definition of $\gamma_5$ for
multiloop calculations, which is most commonly employed in dimensional 
regularization and does not suffer from inconsistencies is~\cite{Veltman}:

\begin{equation}
\gamma_5=i\,\frac{1}{4!}\,\epsilon_{\nu_1\,\nu_2\,\nu_3\,\nu_4}\,\gamma_{\nu_1}\,\gamma_{\nu_2}\,\gamma_{\nu_3}\,\gamma_{\nu_4}
\quad,\quad \nu_i=0,\,1,\,2,\,3
\label{gamma5}
\end{equation}

Of course, $\gamma_5$ as defined in Eq.(\ref{gamma5}) does not anticommute
with the D-dimensional $\gamma_{\mu}$;  an ultimate consequence of
this fact is that Ward identities involving the axial and pseudoscalar
operators, renormalized in this way, are violated. 

To obtain the correctly renormalized pseudoscalar operator, one must
introduce an extra {\em finite} factor, $Z_5$, in addition to
the usual renormalization function $Z^{DR,\overline{MS}}_P$ which only
contains poles in $\epsilon$. We set:  
\begin{equation}
{\cal O}_P=Z_5(g)\,Z^{DR,\overline{MS}}_P\,{\cal O}_{P\,{\rm o}}
\label{pseudoCurrent2}
\end{equation}
$Z_5$ is defined by the requirement that the scalar and pseudoscalar 
renormalized Green's functions coincide: 
\begin{equation}
Z_5\equiv\frac{G_S^{\overline{MS}}\,\gamma_5}{G_P^{\overline{MS}}}
\label{Z5def}
\end{equation}
The value of $Z_5$, calculated in Ref. \cite{Larin}, is gauge
independent; it equals:
\begin{equation}
Z_5(g)=1-\frac{g^2}{16\pi^2}\,(8\,c_F)+\frac{g^4}{(16\pi^2)^2}\,
\left(\frac{2}{9}\,c_F\,N_c+\frac{4}{9}\,c_F\,N_f\right)+{\cal O}(g^6)
\label{Z5}
\end{equation}
$Z_P^{L,\overline{MS}}$ can now be obtained by:

\begin{equation}
Z_P^{L,\overline{MS}} = Z_P^{L,RI^{\prime}} / \left( C_S\,Z_5 \right)
\label{PseudoConversion}
\end{equation}

Similarly, one can convert the $RI^{\prime}$ renormalized Green's functions,
$G_\Gamma^{RI^{\prime}}$, to their $\overline{MS}$ counterparts, through:

\begin{equation}
\frac{G^{RI^{\prime}}_S}{G^{\overline{MS}}_S} = C_{\psi}\,C_S \qquad,\qquad
\frac{G^{RI^{\prime}}_P}{G^{\overline{MS}}_P} = C_{\psi}\,C_S\,Z_5 
\label{RenormGreen}
\end{equation}
(In Eqs.(\ref{PseudoConversion}, \ref{RenormGreen}) it is understood
that powers of $g_{RI^{\prime}},\ \alpha_{RI^{\prime}}$, implicit in
$RI^{\prime}$ quantities, must also be converted to 
$g_{\overline{MS}},\ \alpha_{\overline{MS}}$, respectively, using
Eqs.(\ref{aConversion}, \ref{alphaConversion}). )

\subsection{Fermion Mass Renormalization}

As a by-product of this work, one can evaluate the fermion multiplicative
mass renormalization, $Z_m$, which is directly related to the scalar 
flavor singlet operator. 
This operator differs from the ones
considered thus far, in that it receives also an additive
renormalization, since it has a nonzero perturbative vacuum
expectation value; thus, it mixes with the identity at the 
quantum level. 
Once its vacuum expectation value is subtracted, the
resulting operator is multiplicative renormalizable. The
renormalization is then simply given by only connected
diagrams of the original operator (Figs. 3, 4 and 5); all disconnected
diagrams are easily shown to cancel out.

The perturbative vacuum expectation value\footnote{For a tree level
computation of this quantity, see Ref.~\cite{Sint}.} is of course a power
divergent quantity, and it cannot be expected to
approach well the value of the corresponding disconnected matrix
elements in numerical simulations. Fortunately, this quantity is not
needed for multiplicative renormalization, as mentioned above. 
However, as regards simulations, one should bear in mind that disconnected
parts must be evaluated and subtracted from matrix elements, before
the latter can be renormalized.

Let us express the fermion self energy in the following way:
\begin{equation}
\Sigma_{\psi}^{L,RI^{\prime}}=i\slashed q\,\Sigma_{\rm odd}(qa_{_{\rm L}},m_{\rm o}a_{_{\rm L}},g_{\rm o})+\openone \cdot \frac{1}{a_{_{\rm L}}}\,\Sigma_{\rm even}(qa_{_{\rm L}},m_{\rm o}a_{_{\rm L}},g_{\rm o})
\label{Sigmapsi}
\end{equation}
where $\Sigma_{\rm odd}=1+{\cal O}(g^2_{\rm o})$ 
and $\Sigma_{\rm even}=m_{\rm o}\,a_{_{L}}+{\cal O}(g^2_{\rm o})$. 
Terms like $\sum_{\mu} \,q_{\mu}^3\,\gamma_{\mu}/q^2$, 
though a priori allowed by hypercubic symmetry, are
eventually seen to cancel, as expected by Lorentz invariance.

For generic values of $m_{\rm o}$, the even part of $\Sigma_{\psi}^{L,RI^{\prime}}$
is power divergent; in order to achieve a finite renormalized mass,
$m_r$, the values of the Lagrangian mass $m_{\rm o}$ must be near
a critical value, $m_{\rm cr}$, at which $\Sigma_{\rm even}$ vanishes:
$\Sigma_{\rm even}(qa_{_{\rm L}},m_{\rm cr}a_{_{\rm L}},g_{\rm o})=0+{\cal O}(q^2a_{_{\rm L}}^2)$.
That is, $m_{\rm cr}$ is required to satisfy:
\begin{equation}
\Sigma_{\rm even}(0,m_{\rm cr}a_{_{\rm L}},g_{\rm o})=0
\label{McrCondition} 
\end{equation}
This is a recursive equation which can be solved for $m_{\rm cr}$
order-by-order in perturbation theory. Its value is known to two loops
for Wilson fermions: \cite{Follana} (confirmed independently in
\cite{Rago}), and for clover fermions: \cite{Proestos} (with Wilson
gluons), \cite{Apostolos} (with Symanzik gluons). Only the 1-loop value of 
$m_{\rm cr}$ enters the present calculation.  

We can perform a Taylor expansion with respect to the bare 
mass\footnote{Note that $m_{\rm cr}$ (and, consequently, $m_{\rm o}$)
is power divergent in $a_{_{\rm L}}$ since its calculation
contains no other dimensional quantities; $m_B$, on the other hand 
is at most logarithmically divergent in $a_{_{\rm L}}$.}, 
$m_B\equiv m_{\rm o}-m_{\rm cr}$, for both $\Sigma_{\rm odd}$ 
and $\Sigma_{\rm even}$:
\begin{equation}
\Sigma_{\rm odd}(qa_{_{\rm L}},m_{\rm o}a_{_{\rm L}},g_{\rm o})=
\bigg[\Sigma_{\rm odd}(qa_{_{\rm L}},m_{\rm o}a_{_{\rm L}},g_{\rm o})\bigg]_{m_{\rm o}=m_{\rm cr}}+{\cal O}\Big(m_B\,a_{_{\rm L}}\Big)
\label{SigmaOdd}
\end{equation}

\begin{eqnarray}
\frac{1}{a_{_{\rm L}}}\Sigma_{\rm even}(qa_{_{\rm L}},m_{\rm o}a_{_{\rm L}},g_{\rm o})&=&
\frac{1}{a_{_{\rm L}}}\Bigg[\Sigma_{\rm even}(qa_{_{\rm L}},m_{\rm o}a_{_{\rm L}},g_{\rm o})\Bigg]_{m_{\rm o}=m_{\rm cr}} \nonumber \\
&+& m_B\left[\frac{\partial}{\partial\,(m_{\rm o}a_{_{\rm L}})}\Sigma_{\rm even}(qa_{_{\rm L}},m_{\rm o}a_{_{\rm L}},g_{\rm o})\right]_{m_{\rm o}=m_{\rm cr}}  + {\cal O}(a_{_{\rm L}})
\label{SigmaEven}
\end{eqnarray}
Note that when $m_{\rm o}=m_{\rm cr}$, the first term on the r.h.s. of
Eq.(\ref{SigmaEven}) vanishes in the limit $a_{_{\rm L}}\to 0$, by
virtue of Eq.(\ref{McrCondition}).

Having in mind that, in calculating $\Sigma_{\psi}^{L,RI^{\prime}}$, 
one is interested in the limit $a_{_{\rm L}}\to 0$, the fermion self 
energy takes the form:
\begin{eqnarray}
\Sigma_{\psi}^{L,RI^{\prime}}&=&i\slashed q\,\Bigg[\Sigma_{\rm odd}(qa_{_{\rm L}},m_{\rm o}a_{_{\rm L}},g_{\rm o})\Bigg]_{m_{\rm o}=m_{\rm cr}} \nonumber \\
&+& \openone \cdot m_B\left[\frac{\partial}{\partial\,(m_{\rm o}a_{_{\rm L}})}\Sigma_{\rm even}(qa_{_{\rm L}},m_{\rm o}a_{_{\rm L}},g_{\rm o})\right]_{m_{\rm o}=m_{\rm cr}} 
\label{SigmaPsiTaylor}
\end{eqnarray}
The renormalized fermion mass is now defined by:
\begin{equation}
m_r = \left(Z_m^{L,RI^{\prime}}\right)^{-1}\,m_B
\label{Zm}
\end{equation}
The renormalization condition for $Z_{\psi}^{L,RI^{\prime}}$ (Eq.(\ref{ZpsiRule})) 
for nonzero $m_r$, reads:
\begin{equation}
\lim_{a_{_{\rm L}}\to 0}\bigg[Z_{\psi}^{L,RI^{\prime}}\,\Sigma_{\psi}^{L,RI^{\prime}}-\left(i\slashed q + m_r\right)\bigg]_{q^2=\bar{\mu}^2}=0
\label{ZpsiRule2}
\end{equation}

By combining Eqs.(\ref{Zm}) and (\ref{ZpsiRule2}) we find the renormalization
condition for $Z_m^{L,RI^{\prime}}$:
\begin{equation}
\lim_{a_{_{\rm L}}\to 0}\Bigg[Z_{\psi}^{L,RI^{\prime}}\,Z_m^{L,RI^{\prime}}\,\bigg[\frac{\partial}{\partial\,(m_{\rm o}a_{_{\rm L}})}\Sigma_{\rm even}(qa_{_{\rm L}},m_{\rm o}a_{_{\rm L}},g_{\rm o})\bigg]_{m_{\rm o}=m_{\rm cr}}\Bigg]_{q^2=\bar{\mu}^2}=1
\label{ZmRule}
\end{equation}
We stress again that, even though the Lagrangian mass $m_{\rm o}$
may take arbitrary values, the renormalization condition involves 
only $m_{\rm o}\to m_{\rm cr}$.

In order to establish a relation between $Z_m^{L,RI^{\prime}}$ and 
$Z_{S,\,singlet}^{L,RI^{\prime}}$, note that Eq.(\ref{ZmRule})
coincides with Eq.(\ref{ZXrulesa}) if  
$\partial/\partial\,(m_{\rm o}a_{_{\rm L}})\Sigma_{\rm even}=
\Sigma_{S,\,singlet}^{L,RI^{\prime}}$. Indeed, the equality between 
$\partial/\partial\,(m_{\rm o}a_{_{\rm L}})\Sigma_{\rm even}$
and $\Sigma_{S,\,singlet}^{L,RI^{\prime}}$ holds diagram by diagram in
perturbation theory, noting that:
\begin{itemize}
\item The tree level value equals 1, in both cases 
\item The effect of inserting the scalar operator on a given fermion
propagator of any self-energy Feynman diagram is equivalent to taking the
negative partial derivative $-\partial/\partial\,(m_{\rm o}a_{_{\rm L}})$
of that propagator
\item Combinatorial factors agree 
\item There is an extra minus sign in the geometric series summation
of 1PI diagrams leading to the fermion self-energy 
\end{itemize}

Once all of the above statements are taken into account, one comes
to the conclusion that:
\begin{equation}
Z_m^{L,RI^{\prime}}=Z_{S,\,singlet}^{L,RI^{\prime}}
\label{ZmZS}
\end{equation}

Given that $m_{\rm cr}\,a_{_{\rm L}}={\cal O}(g^2_{\rm o})$, all two-loop
calculations can be performed with strictly massless fermion propagators,
provided that appropriate fermion mass counterterms are introduced on
one-loop diagrams.


\section{Computation and Results}
\label{Results}

The Feynman diagrams relevant to the fermion self-energy
$\Sigma^L_{\psi}(q,a_{_{\rm L}})$, at 1- and 2-loop level, are shown
in Figs.1 and 2, respectively; those relevant to
$\Sigma^L_S(q a_{_{\rm L}})$, $\Sigma^L_P(q a_{_{\rm L}})$ are shown
in Figs.3 and 4. 

For flavor singlet bilinears, there are 4 extra
diagrams, in addition to those of Fig.4, shown in Fig.5; 
in these diagrams, the
operator insertion occurs inside a closed fermion loop.

The evaluation and algebraic manipulation of
Feynman diagrams, leading to a code for numerical loop integration, is
performed automatically using our software for Lattice Perturbation
Theory, written in Mathematica. 

The most laborious aspect of the procedure is the extraction of
the dependence on the external momentum $q$. This is a delicate task at two
loops; for this purpose, we cast algebraic expressions (typically
involving thousands of summands) into terms which can be naively Taylor
expanded in $q$ to the required order, plus a
smaller set of terms containing superficial divergences and/or
subdivergences. The latter can be evaluated by an extension of the method of
Ref.~\cite{KNS} to 2 loops; this entails analytical continuation to
$D>4$ dimensions, and splitting each expression into a UV-finite part
(which can thus be calculated in the continuum, using the methods of
Ref.~\cite{Chetyrkin}), and a part which is polynomial in $q$.
A primitive set of divergent lattice integrals involving gluon
propagators, which can be obtained in this manner,
can be found in Ref.~\cite{Luscher}.

Some of the diagrams contributing to $\Sigma^L_{\psi}(q,a_{_{\rm L}})$,
$\Sigma^L_S(q a_{_{\rm L}})$ and $\Sigma^L_P(q a_{_{\rm L}})$ are infrared divergent
when considered separately, and thus must be grouped together in order to give finite
results. Such groups are formed by diagrams (7-11), (12-13), (14-18), (19-20),
(21-23) in Fig.2, diagrams (3-7), (8-9), (10-11,19) in Fig.4 and 
diagrams (1-2), (3-4) in Fig.5. 

In Figures 1-5, ``mirror'' diagrams (those in which the
direction of the external fermion line is reversed) should also be
taken into account. In most cases, these coincide trivially with the
original diagrams; even in the remaining cases,
they can be seen to give equal contribution, by invariance under charge conjugation.

As mentioned before, all calculations should be performed at 
vanishing renormalized mass; this can be achieved by working with
massless fermion propagators, provided an appropriate fermion mass
counterterm is introduced (diagram 23 in Fig.2 and diagram 11 in
Fig.4). 

All two-loop diagrams have been calculated in the
bare Feynman gauge ($\alpha_{\rm o}=1$). One-loop 
diagrams have been calculated 
for generic values of $\alpha_{\rm o}$; this allows 
us to convert our two-loop results
to the renormalized Feynman gauge ($\alpha_{RI^{\prime}}=1$ 
or $\alpha_{\overline{MS}}=1$). 

Numerical loop integration was carried out by our ``integrator''
program, a {\em metacode} written in Mathematica, for 
converting lengthy integrands into efficient Fortran code.
Two-loop numerical integrals were evaluated for 
lattices of size up to $L=40$; the results were then 
extrapolated to $L\rightarrow\infty$.
Extrapolation is the only source of systematic error; this error can
be estimated quite accurately (see, e.g. Ref.~\cite{PST}), given that
$L$-dependence of results can only span a restricted set of functional forms.

\newpage
\begin{center}
\psfig{figure=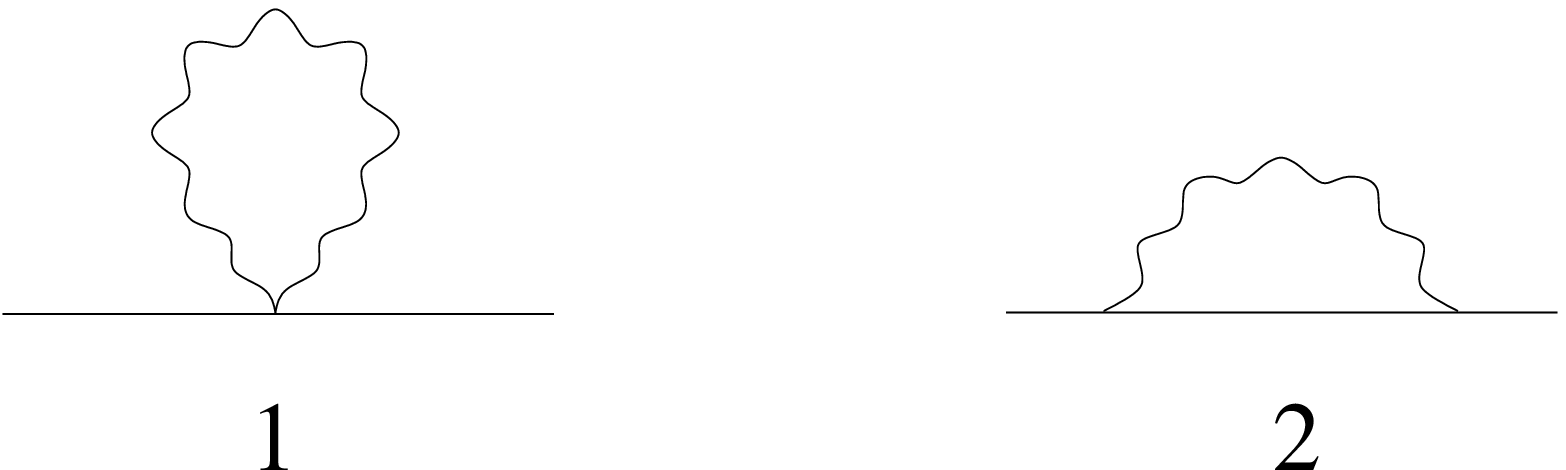,height=1.8truecm}
\vskip 1mm
{\small FIG. 1. One-loop diagrams contributing to $Z_{\psi}$.\\
A wavy (solid) line represents gluons (fermions).}
\end{center}

\begin{center}
\psfig{figure=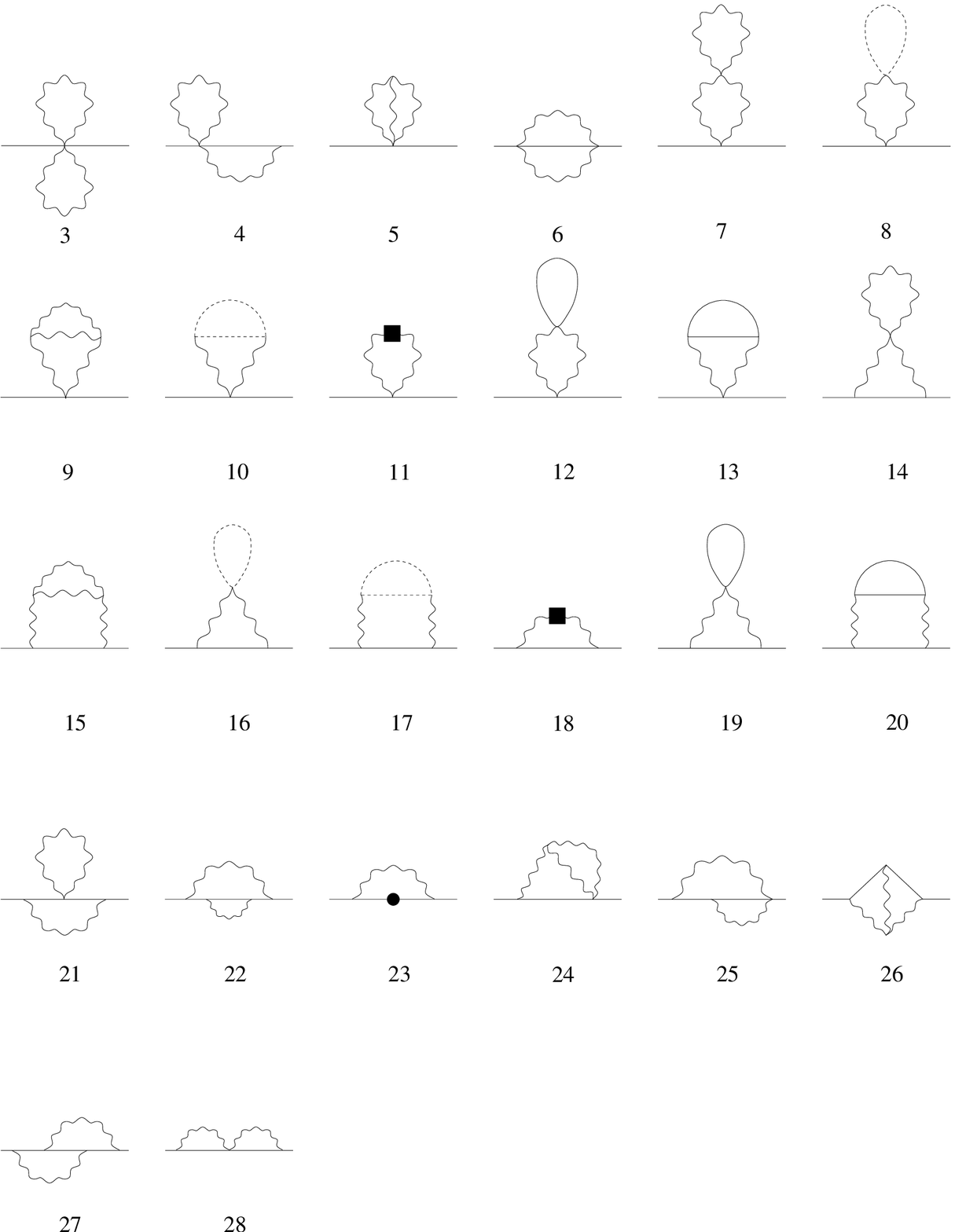,scale=0.72}
\vskip 1mm
{\small FIG. 2. Two-loop diagrams contributing to $Z_{\psi}$.
Wavy (solid, dotted) lines represent gluons (fermions,
ghosts). Solid boxes denote vertices stemming from the measure part of the
action; a solid circle is a fermion mass counterterm.}
\end{center}

\newpage
\begin{center}
\psfig{figure=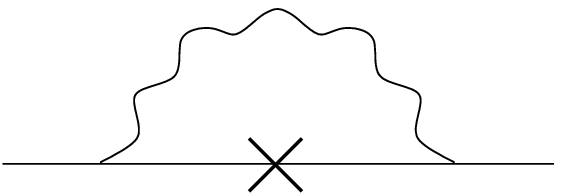,scale=0.40}
\vskip 1mm
{\small FIG. 3. One-loop diagram contributing to $Z_{S}$ and $Z_{P}$.
A wavy (solid) line represents gluons (fermions). A cross denotes the
Dirac matrices $\openone$ (scalar) and $\gamma_{5}$ (pseudoscalar).}
\end{center}

\begin{center}
\psfig{figure=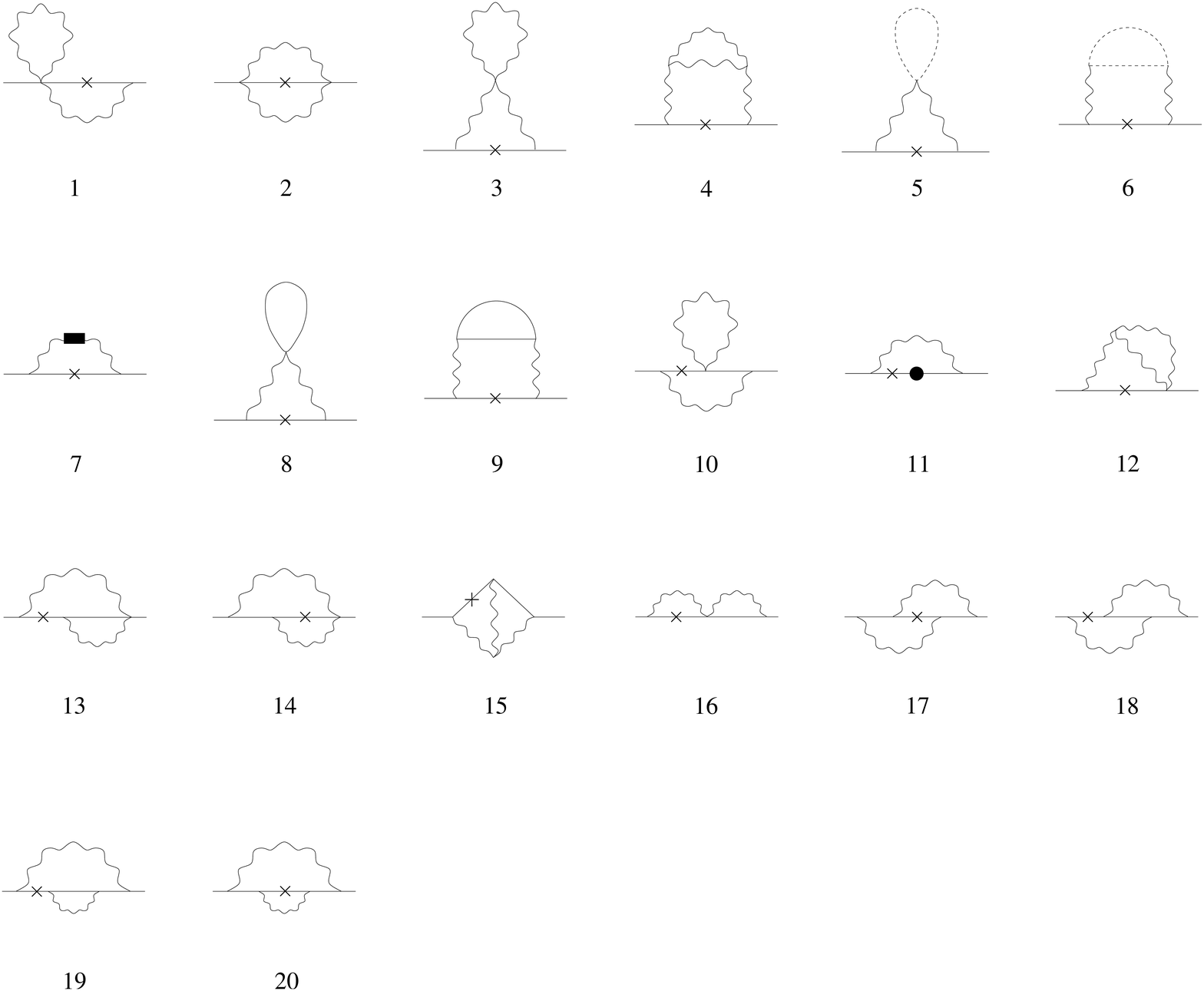,scale=0.35}
\vskip 2mm
{\small FIG. 4. Two-loop diagrams contributing to $Z_{S}$ and $Z_{P}$.
Wavy (solid, dotted) lines represent gluons (fermions, ghosts). A solid box 
denotes a vertex from the measure part of the action; a solid circle 
is a mass counterterm; crosses denote the matrices $\openone$ 
(scalar) and $\gamma_{5}$ (pseudoscalar).}
\end{center}

\begin{center}
\psfig{figure=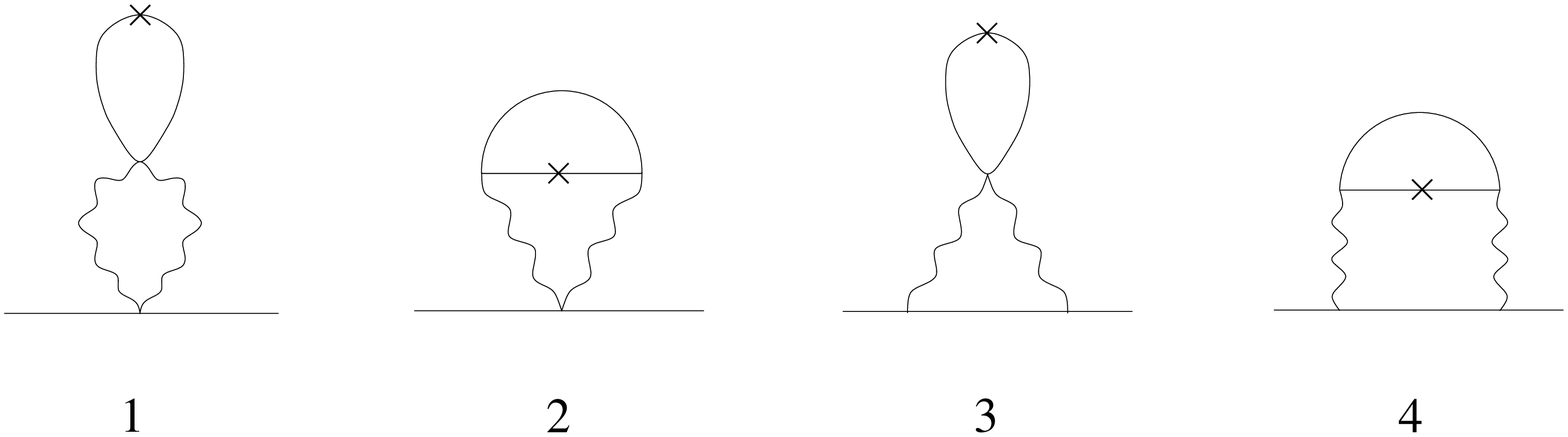,height=3.0truecm}
\vskip 1mm
{\small FIG. 5. Extra two-loop diagrams contributing to $Z_{S,\,singlet}$.
A cross denotes an insertion of a flavor singlet operator.
Wavy (solid) lines represent gluons (fermions)}.
\end{center}

\subsection{One-loop results} 

1-loop results for $Z_{\psi}^{L,RI^{\prime}}$, 
$Z_S^{L,RI^{\prime}}$ and $Z_P^{L,RI^{\prime}}$ are presented below in 
a generic gauge. The errors result from the $L\to\infty$ extrapolation. 

\begin{eqnarray}
Z_{\psi}^{L,RI^{\prime}} = 1 + \frac{g_\circ^2}{16\pi^2}\,c_F &\bigg{[}&\left ( \ln(a_{_{\rm L}}^2 \bar{\mu}^2)-4.792009570(1)\right )\,\alpha_{\rm o} + 16.644413858(5) \nonumber \\
&&\,\, - 2.248868528(3)\,c_{{\rm SW}} - 1.397267102(5)\,c_{{\rm SW}}^2 \bigg{]} \label{Zpsi1loopRI} \\
Z_S^{L,RI^{\prime}}= 1 + \frac{g_\circ^2}{16\pi^2}\,c_F &\bigg{[}&3\,\ln(a_{_{\rm L}}^2 \bar{\mu}^2)- \alpha_{\rm o} - 16.9524103(1) \nonumber \\
&& \,\, - 7.7379159(3)\,c_{{\rm SW}} + 1.38038065(4)\,c_{{\rm SW}}^2 \bigg{]} \label{ZS1loopRI} \\
Z_P^{L,RI^{\prime}}= 1 + \frac{g_\circ^2}{16\pi^2}\,c_F&\bigg{[}&3\,\ln(a_{_{\rm L}}^2 \bar{\mu}^2)- \alpha_{\rm o} - 26.5954414(1) \nonumber \\
&&\,\, + 2.248868528(3)\,c_{{\rm SW}} - 2.03601561(4)\,c_{{\rm SW}}^2 \bigg{]} \label{ZP1loopRI}
\end{eqnarray}

The corresponding quantities in the $\overline{MS}$ scheme are:

\begin{eqnarray}
Z_{\psi}^{L,\overline{MS}} = 1 + \frac{g_\circ^2}{16\pi^2}\,c_F &\bigg{[}& \left ( \ln(a_{_{\rm L}}^2 \bar{\mu}^2) - 3.792009570(1) \right )\,\alpha_{\rm o} + 16.644413858(5) \nonumber \\
&& \,\, - 2.248868528(3)\,c_{{\rm SW}} - 1.397267102(5)\,c_{{\rm SW}}^2  \bigg{]} \label{Zpsi1loopMS} \\
Z_S^{L,\overline{MS}} = 1 + \frac{g_\circ^2}{16\pi^2}\,c_F &\bigg{[}& 3\,\ln(a_{_{\rm L}}^2 \bar{\mu}^2) - 12.9524103(1) \nonumber \\
&& \,\, - 7.7379159(3)\,c_{{\rm SW}} + 1.38038065(4)\,c_{{\rm SW}}^2 \bigg{]} \label{ZS1loopMS} \\
Z_P^{L,\overline{MS}} = 1 + \frac{g_\circ^2}{16\pi^2}\,c_F &\bigg{[}& 3\,\ln(a_{_{\rm L}}^2 \bar{\mu}^2) -14.5954414(1) \nonumber \\
&& \,\, + 2.248868528(3)\,c_{{\rm SW}} - 2.03601561(4)\,c_{{\rm SW}}^2 \bigg{]} \label{ZP1loopMS}
\end{eqnarray}

Our results confirm
the existing results found in the literature~\cite{Capitani} (note,
however, a difference in $Z_P^{L,\overline{MS}}$; this is entirely due
to the factor $Z_5$ in Eq.(\ref{PseudoConversion})\,). 

\newpage
\subsection{Two-loop results} 

The evaluation of all Feynman diagrams in Figs.1-5 leads directly to
the corresponding bare Green's functions $\Sigma^L_{\psi}$,
$\Sigma^L_{S}$ and $\Sigma^L_{P}$. These, in turn, can be converted to
the corresponding renormalization functions $Z_{\psi}^{L,Y}$,
$Z_{S}^{L,Y}$ and $Z_{P}^{L,Y}$ ($Y=RI^{\prime}$ or $\overline{MS}$),
via Eqs.(\ref{ZpsiRule}), (\ref{ZXrulesa}) and (\ref{ZXrulesb}). To
this end, we need the following one-loop expression for 
$Z_A^{L,Y}$ (recall that $Z_{\alpha}=1$ to this order): 

\begin{eqnarray}
Z_A^{L,RI^{\prime}}&=& Z_A^{L,\overline{MS}} + {\cal O}\left(g_{\rm o}^4\right) \nonumber \\
&=&1 + \frac{g_\circ^2}{16\pi^2} \bigg[\ln\left(a_{_{\rm L}}^2\bar{\mu}^2\right)\left(\frac{2}{3}\,N_f-\frac{5}{3}\,N_c \right) \nonumber \\
&& \hskip 1.15cm + N_f\,\left(-2.168501047(1) + 0.7969452308(4)\,c_{\rm SW} -4.7126914428(1)\,c_{\rm SW}^2\right) \nonumber \\ 
&& \hskip 1.15cm  + 39.47841760436(1)\,c_F +1.94017130069(1)\,N_c\bigg] + {\cal O}\left(g_{\rm o}^4\right) 
\label{ZA1loop}
\end{eqnarray}

To express our results in terms of the renormalized coupling constant, we also need the one-loop expression 
for $Z_g^{L,Y}$:
\begin{eqnarray}
Z_g^{L,RI^{\prime}}&=& Z_g^{L,\overline{MS}} + {\cal O}\left(g_{\rm o}^4\right) \nonumber \\
&=&1 + \frac{g_\circ^2}{16\pi^2} \bigg[\ln\left(a_{_{\rm L}}^2\bar{\mu}^2\right)\left(-\frac{1}{3}\,N_f+\frac{11}{6}\,N_c \right)  \nonumber \\
&& \hskip 1.15cm + N_f\,\left(0.5286949677(5) -0.3984726154(2)\,c_{\rm SW} + 2.35634572140(7)\,c_{\rm SW}^2\right) \nonumber \\ 
&& \hskip 1.15cm -19.73920880218(1)\,c_F -3.54958342046(1)\,N_c\bigg] + {\cal O}\left(g_{\rm o}^4\right)
\label{Zg1loop}
\end{eqnarray}
Eqs.(\ref{ZA1loop}, \ref{Zg1loop}) are in agreement with older references (see, e.g., Ref.~\cite{Bode}).

We present below $Z_{\psi}^{L,RI^{\prime}}$, $Z_{S}^{L,RI^{\prime}}$
and $Z_{P}^{L,RI^{\prime}}$ to two loops in the renormalized Feynman
gauge $\alpha_{RI^{\prime}}=1$; we also present the $\overline{MS}$ analogues
$Z_{\psi}^{L,\overline{MS}}$, $Z_{S}^{L,\overline{MS}}$
and $Z_{P}^{L,\overline{MS}}$ in the gauge
$\alpha_{\overline{MS}}=1$. For conciseness, we omit the bare Green's
functions; it is a straightforward exercise to recover these from the  
corresponding $Z$'s. 

\newpage
\begin{eqnarray}
Z_{\psi}^{L,RI^{\prime}} = 1 &+& \frac{g_\circ^2}{16\pi^2}\,c_F \bigg{[} \ln(a_{_{\rm L}}^2 \bar{\mu}^2) + 11.852404288(5) - 2.248868528(3)\,c_{{\rm SW}} - 1.397267102(5)\,c_{{\rm SW}}^2 \bigg{]} \nonumber \\
&+& \frac{g_\circ^4}{(16\pi^2)^2}\,c_F \Bigg{[} \ln^2(a_{_{\rm L}}^2 \bar{\mu}^2) \left ( \frac{1}{2} c_F + \frac{2}{3} N_f -\frac{8}{3} N_c \right ) \nonumber \\
&&\quad\quad\,\, + \ln(a_{_{\rm L}}^2 \bar{\mu}^2) \Big{(}-6.36317446(8)\,N_f + 0.79694523(2)\,N_f\,c_{{\rm SW}} \nonumber \\
&& \hskip 3.2cm - 4.712691443(4)\,N_f\,c_{{\rm SW}}^2 \nonumber \\
&& \hskip 3.2cm +49.83082185(5)\,c_F - 2.24886861(7)\,c_F\,c_{{\rm SW}} \nonumber \\
&& \hskip 3.2cm - 1.39726705(1)\,c_F\,c_{{\rm SW}}^2  +29.03029398(4)\,N_c\Big{)} \nonumber \\
&&\quad\quad\,\, + N_f\,\Big{(}-7.838(2) + 1.153(1)\,c_{{\rm SW}} + 3.202(3)\,c_{{\rm SW}}^2 \nonumber \\
&& \hskip 2.2cm +6.2477(6)\,c_{{\rm SW}}^3 + 4.0232(6)\,c_{{\rm SW}}^4 \Big{)}  \nonumber \\
&&\quad\quad\,\, + c_F\,\Big{(}505.39(1) - 58.210(9)\,c_{{\rm SW}} + 20.405(5)\,c_{{\rm SW}}^2 \nonumber \\
&& \hskip 2.2cm +18.8431(8)\,c_{{\rm SW}}^3 + 4.2793(2)\,c_{{\rm SW}}^4 \Big{)}  \nonumber \\
&&\quad\quad\,\, + N_c\,\Big{(}-20.59(1) - 3.190(5)\,c_{{\rm SW}} -
  23.107(6)\,c_{{\rm SW}}^2  \nonumber \\
&& \hskip 2.2cm -5.7234(5)\,c_{{\rm SW}}^3 -
  0.7938(1)\,c_{{\rm SW}}^4 \Big{)} \Bigg{]} 
\label{Zpsi2loopRI}\\[2.0ex]
Z_{\psi}^{L,\overline{MS}} = 1 &+& \frac{g_\circ^2}{16\pi^2}\,c_F\,\bigg{[} \ln(a_{_{\rm L}}^2 \bar{\mu}^2) +12.852404288(5) 
-2.248868528(3)\,c_{{\rm SW}} -1.397267102(5)\,c_{{\rm SW}}^2 \bigg{]} \nonumber \\
&+& \frac{g_\circ^4}{(16\pi^2)^2}\,c_F\,\Bigg{[} \ln^2(a_{_{\rm L}}^2 \bar{\mu}^2) \left ( \frac{2}{3} N_f +\frac{1}{2}\,c_F - \frac{8}{3}\,N_c \right )  \nonumber \\
&&\quad\quad\,\, + \ln(a_{_{\rm L}}^2 \bar{\mu}^2) \Big{(}-4.58539668(8)\,N_f + 0.79694523(2)\,N_f\,c_{{\rm SW}} \nonumber \\
&& \hskip 3.2cm -4.712691443(4)\,N_f\,c_{{\rm SW}}^2 \nonumber \\
&& \hskip 3.2cm +50.83082185(5)\,c_F -2.24886861(7)\,c_F\,c_{{\rm SW}} \nonumber \\
&& \hskip 3.2cm -1.39726705(1)\,c_F\,c_{{\rm SW}}^2  +21.91918287(4)\,Nc \Big{)} \nonumber \\
&&\quad\quad\,\, + N_f\,\Big{(}-15.970(2) + 1.950(1)\,c_{{\rm SW}} -1.510(3)\,c_{{\rm SW}}^2 \nonumber \\
&& \hskip 2.2cm +6.2477(6)\,c_{{\rm SW}}^3 +4.0232(6)\,c_{{\rm SW}}^4 \Big{)}  \nonumber \\
&&\quad\quad\,\, + c_F\,\Big{(}556.10(1) -60.459(9)\,c_{{\rm SW}} +19.007(5)\,c_{{\rm SW}}^2 \nonumber \\
&& \hskip 2.2cm +18.8431(8)\,c_{{\rm SW}}^3 +4.2793(2)\,c_{{\rm SW}}^4 \Big{)} \nonumber \\
&&\quad\quad\,\, + N_c\,\Big(13.68(1) -3.190(5)\,c_{{\rm SW}} -23.107(6)\,c_{{\rm SW}}^2 \nonumber \\
&&\quad\quad\,\, -5.7234(5)\,c_{{\rm SW}}^3 -0.7938(1)\,c_{{\rm SW}}^4\Big)  \Bigg{]} 
\label{Zpsi2loopMS}
\end{eqnarray}
\newpage
\begin{eqnarray}
Z_S^{L,RI^{\prime}} = 1 &+& \frac{g_\circ^2}{16\pi^2}\,c_F \bigg{[} 3\,\ln(a_{_{\rm L}}^2 \bar{\mu}^2) -17.9524103(1) 
-7.7379159(3)\,c_{{\rm SW}} + 1.38038065(4)\,c_{{\rm SW}}^2 \bigg{]} \nonumber \\
&+& \frac{g_\circ^4}{(16\pi^2)^2}\,c_F \Bigg{[} \ln^2(a_{_{\rm L}}^2 \bar{\mu}^2) \left ( \frac{9}{2} c_F + N_f -\frac{11}{2} N_c \right ) \nonumber \\
&&\quad\quad\,\, + \ln(a_{_{\rm L}}^2 \bar{\mu}^2) \Big{(}-8.1721694(5)\,N_f + 2.3908354(3)\,N_f\,c_{{\rm SW}} \nonumber \\
&& \hskip 3.2cm -14.13807433(4)\,N_f\,c_{{\rm SW}}^2 \nonumber \\
&& \hskip 3.2cm +66.0780218(9)\,c_F - 23.213749(2)\,c_F\,c_{{\rm SW}} \nonumber \\ 
&& \hskip 3.2cm + 4.1411425(3)\,c_F\,c_{{\rm SW}}^2 +55.7975008(9)\,N_c\Big{)} \nonumber \\
&&\quad\quad\,\, + N_f\,\Big{(}24.003(3) + 11.878(5)\,c_{{\rm SW}} + 25.59(1)\,c_{{\rm SW}}^2 \nonumber \\
&& \hskip 2.2cm + 22.078(3)\,c_{{\rm SW}}^3 -6.1807(8)\,c_{{\rm SW}}^4 \Big{)}  \nonumber \\
&&\quad\quad\,\, + c_F\,\Big{(}-602.35(6) + 91.07(7)\,c_{{\rm SW}} + 51.15(5)\,c_{{\rm SW}}^2  \nonumber \\
&& \hskip 2.2cm -27.759(4)\,c_{{\rm SW}}^3 -2.688(1)\,c_{{\rm SW}}^4 \Big{)}  \nonumber \\
&&\quad\quad\,\, + N_c\,\Big{(}-38.16(4) -132.40(5)\,c_{{\rm SW}} -4.04(3)\,c_{{\rm SW}}^2 \nonumber \\
&& \hskip 2.2cm +12.576(3)\,c_{{\rm SW}}^3 +1.0175(8)\,c_{{\rm SW}}^4 \Big{)} \Bigg{]} 
\label{ZS2loopRI}\\[2.0ex]
Z_S^{L,\overline{MS}} = 1 &+& \frac{g_\circ^2}{16\pi^2}\,c_F\,\bigg{[} 3\,\ln(a_{_{\rm L}}^2 \bar{\mu}^2) -12.9524103(1) 
-7.7379159(3)\,c_{{\rm SW}} + 1.38038065(4)\,c_{{\rm SW}}^2 \bigg{]} \nonumber \\
&+& \frac{g_\circ^4}{(16\pi^2)^2}\,c_F\,\Bigg{[} \ln^2(a_{_{\rm L}}^2 \bar{\mu}^2) \left( N_f + \frac{9}{2}\,c_F - \frac{11}{2}\,N_c \right)  \nonumber \\
&&\quad\quad\,\, + \ln(a_{_{\rm L}}^2 \bar{\mu}^2) \Big{(}-4.8388361(5)\,N_f + 2.3908354(3)\,N_f\,c_{{\rm SW}} \nonumber \\
&& \hskip 3.2cm -14.13807433(4)\,N_f\,c_{{\rm SW}}^2 \nonumber \\
&& \hskip 3.2cm +81.0780218(9)\,c_F -23.213749(2)\,c_F\,c_{{\rm SW}}\nonumber \\
&& \hskip 3.2cm +4.1411425(3)\,c_F\,c_{{\rm SW}}^2 +37.4641674(9)\,N_c \Big{)} \nonumber \\
&&\quad\quad\,\, + N_f\,\Big{(}10.688(3)+15.863(5)\,c_{{\rm SW}}+2.02(1)\,c_{{\rm SW}}^2 \nonumber \\
&& \hskip 2.2cm +22.078(3)\,c_{{\rm SW}}^3 -6.1807(8)\,c_{{\rm SW}}^4 \Big{)} \nonumber \\
&&\quad\quad\,\, + c_F\,\Big(-462.67(6)+52.38(7)\,c_{{\rm SW}} +58.05(5)\,c_{{\rm SW}}^2 \nonumber \\
&& \hskip 2.2cm -27.759(4)\,c_{{\rm SW}}^3 -2.688(1)\,c_{{\rm SW}}^4\Big)   \nonumber \\
&&\quad\quad\,\, + N_c\,\Big(36.93(4)-132.40(5)\,c_{{\rm SW}}-4.04(3)\,c_{{\rm SW}}^2 \nonumber \\
&& \hskip 2.2cm +12.576(3)\,c_{{\rm SW}}^3+1.0175(8)\,c_{{\rm SW}}^4 \Big) \Bigg{]}
\label{ZS2loopMS}
\end{eqnarray}
\newpage
\begin{eqnarray}
Z_P^{L,RI^{\prime}} = 1 &+& \frac{g_\circ^2}{16\pi^2}\,c_F \bigg{[} 3\,\ln(a_{_{\rm L}}^2 \bar{\mu}^2) -27.5954414(1)
+2.248868528(3)\,c_{{\rm SW}} -2.03601561(4)\,c_{{\rm SW}}^2 \bigg{]} \nonumber \\
&+& \frac{g_\circ^4}{(16\pi^2)^2}\,c_F \Bigg{[} \ln^2(a_{_{\rm L}}^2 \bar{\mu}^2) \left ( \frac{9}{2} c_F + N_f -\frac{11}{2} N_c \right ) \nonumber \\
&&\quad\quad\,\, + \ln(a_{_{\rm L}}^2 \bar{\mu}^2) \Big{(}-8.1721694(4)\,N_f + 2.39083540(6)\,N_f\,c_{{\rm SW}} \nonumber \\
&& \hskip 3.2cm-14.13807433(4)\,N_f\,c_{{\rm SW}}^2 \nonumber \\
&& \hskip 3.2cm +37.1489292(7)\,c_F +6.746606(1)\,c_F\,c_{{\rm SW}} \nonumber \\
&& \hskip 3.2cm-6.1080465(3)\,c_F\,c_{{\rm SW}}^2 +55.7975008(7)\,N_c\Big{)} \nonumber \\
&&\quad\quad\,\, + N_f\,\Big{(}38.231(3) -7.672(5)\,c_{{\rm SW}} + 55.32(1)\,c_{{\rm SW}}^2 \nonumber \\
&& \hskip 2.2cm -7.049(3)\,c_{{\rm SW}}^3 +4.7469(8)\,c_{{\rm SW}}^4 \Big{)}  \nonumber \\
&&\quad\quad\,\, + c_F\,\Big{(}-876.98(4) + 84.52(2)\,c_{{\rm SW}} +
  38.65(4)\,c_{{\rm SW}}^2  \nonumber \\
&& \hskip 2.2cm + 19.974(3)\,c_{{\rm SW}}^3 + 2.873(1)\,c_{{\rm SW}}^4 \Big{)}  \nonumber \\
&&\quad\quad\,\, + N_c\,\Big{(}-104.35(3) -38.06(2)\,c_{{\rm SW}} -14.57(3)\,c_{{\rm SW}}^2 \nonumber \\
&& \hskip 2.2cm -4.429(2)\,c_{{\rm SW}}^3 -1.2898(7)\,c_{{\rm SW}}^4 \Big{)} \Bigg{]} 
\label{ZP2loopRI}\\[2.0ex]
Z_P^{L,\overline{MS}} = 1 &+& \frac{g_\circ^2}{16\pi^2}\,c_F\,\bigg{[} 3\,\ln(a_{_{\rm L}}^2 \bar{\mu}^2) -14.5954414(1)
+2.248868528(3)\,c_{{\rm SW}} -2.03601561(4)\,c_{{\rm SW}}^2 \bigg{]} \nonumber \\
&+& \frac{g_\circ^4}{(16\pi^2)^2}\,c_F\,\Bigg{[} \ln^2(a_{_{\rm L}}^2 \bar{\mu}^2) \left( N_f +\frac{9}{2}\,c_F- \frac{11}{2}\,N_c \right)  \nonumber \\
&&\quad\quad\,\, + \ln(a_{_{\rm L}}^2 \bar{\mu}^2) \Big{(}0.4944972(4)\,N_f + 2.39083540(6)\,N_f\,c_{{\rm SW}} \nonumber \\
&& \hskip 3.2cm-14.13807433(4)\,N_f\,c_{{\rm SW}}^2 \nonumber \\
&& \hskip 3.2cm +76.1489292(7)\,c_F +6.746606(1)\,c_F\,c_{{\rm SW}} \nonumber \\
&& \hskip 3.2cm -6.1080465(3)\,c_F\,c_{{\rm SW}}^2 +8.1308341(7)\,N_c \Big{)} \nonumber \\
&&\quad\quad\,\, + N_f\,\Big{(}16.013(3) +2.688(5)\,c_{{\rm SW}} -5.94(1)\,c_{{\rm SW}}^2 \nonumber \\
&& \hskip 2.2cm -7.049(3)\,c_{{\rm SW}}^3 +4.7469(8)\,c_{{\rm SW}}^4 \Big{)} \nonumber \\
&&\quad\quad\,\, + c_F\,\Big{(}-586.45(4)+113.76(2)\,c_{{\rm SW}} +12.18(4)\,c_{{\rm SW}}^2 \nonumber \\
&& \hskip 2.2cm +19.974(3)\,c_{{\rm SW}}^3 +2.873(1)\,c_{{\rm SW}}^4 \Big{)}\nonumber \\
&&\quad\quad\,\, + N_c\,\Big(27.31(3) -38.06(2)\,c_{{\rm SW}} -14.57(3)\,c_{{\rm SW}}^2 \nonumber \\
&& \hskip 2.2cm -4.429(2)\,c_{{\rm SW}}^3 -1.2898(7)\,c_{{\rm SW}}^4 \Big)  \Bigg{]} 
\label{ZP2loopMS}
\end{eqnarray}
\newpage

All expressions reported thus far for $Z_S$ and $Z_P$ refer to flavor
non singlet operators. In the case of $Z_P$, all diagrams
of Fig.5 vanish, so that singlet and non singlet results coincide, just as in
dimensional regularization. For $Z_S$ on the other hand, the above
diagrams give an additional finite contribution:

\begin{eqnarray}
Z_{S,\,\rm singlet}^{L,RI^{\prime}} = Z_S^{L,RI^{\prime}} 
+ \frac{g_\circ^4}{(16\pi^2)^2}\,c_F N_f\,
\Bigl( &-&107.76(1) + 82.27(2)\,c_{{\rm SW}} -29.727(4)\,c_{{\rm SW}}^2
\nonumber \\
&+& 3.4400(7)\,c_{{\rm SW}}^3 + 2.2758(4)\,c_{{\rm SW}}^4\Bigr)
\label{ScalarSingletRI}
\end{eqnarray}
The same extra finite contribution applies also to the $\overline{MS}$ scheme.

Finally, for completeness, and as an additional check on our results, we compute the renormalized Green's
functions (for {\em vanishing} renormalized mass):
\begin{eqnarray}
G_{\psi}^{RI^{\prime}}(\bar{\mu}/q)&\equiv& Z_{\psi}^{L,RI^{\prime}}\,\Sigma_{\rm
odd} \label{Grenormconditionsa} \\
G_{S}^{RI^{\prime}}(\bar{\mu}/q)&\equiv&
Z_{\psi}^{L,RI^{\prime}}\,Z_{S}^{L,RI^{\prime}}\, \Sigma_S^L \label{Grenormconditionsb} \\
G_{P}^{RI^{\prime}}(\bar{\mu}/q)&\equiv&
Z_{\psi}^{L,RI^{\prime}}\,Z_{P}^{L,RI^{\prime}}\, \Sigma_P^L
\label{Grenormconditionsc} 
\end{eqnarray}
Similarly for $\overline{MS}$, taking into account Eq.(\ref{RenormGreen}).

Since these functions are regularization independent, they can be
calculated also using, e.g., dimensional regularization. We have
computed $G_{\psi}$, $G_S$ and $G_P$ in both ways: either starting 
from our Eqs.(\ref{ZA1loop}-\ref{ZP2loopMS}) or using 
renormalization functions from dimensional regularization~\cite{Gracey}. 
In all cases the two ways are in complete agreement. We obtain:
\begin{eqnarray}
G^{RI^{\prime}}_{\psi}=1 &+& \frac{g_{RI^{\prime}}^2}{16\pi^2}\,c_F\ln(\bar{\mu}^2/q^2) \nonumber \\
&+& \frac{g_{RI^{\prime}}^4}{(16\pi^2)^2}\,c_F\bigg[\ln^2(\bar{\mu}^2/q^2)\left(\frac{1}{2}\,c_F+N_c\right) \nonumber \\ 
&&\quad\qquad\,\, + \ln(\bar{\mu}^2/q^2)\left(-\frac{19}{9}\,N_f-\frac{3}{2}\,c_F+\frac{251}{18}\,N_c\right)\bigg]
\label{PsiRenormGreenFnRI}\\
G^{\overline{MS}}_{\psi}=1 &+& \frac{g_{\overline{MS}}^2}{16\pi^2}\,c_F\bigg[\ln(\bar{\mu}^2/q^2)+1\bigg] \nonumber \\
&+& \frac{g_{\overline{MS}}^4}{(16\pi^2)^2}\,c_F\bigg[\ln^2(\bar{\mu}^2/q^2)\left(\frac{1}{2}\,c_F+N_c\right) \nonumber \\  
&& \quad\qquad\,+ \ln(\bar{\mu}^2/q^2)\left(-N_f-\frac{1}{2}\,c_F+\frac{21}{2}\,N_c\right) \nonumber \\
&& \quad\qquad\,+\left(-\frac{7}{4}\,N_f-\frac{5}{8}\,c_F+\left(\frac{143}{8}-6\zeta(3)\right)\,N_c\right)\bigg]
\label{PsiRenormGreenFnMS}
\end{eqnarray}
\begin{eqnarray}
G^{RI^{\prime}}_{S}=1 + \frac{g_{RI^{\prime}}^2}{16\pi^2}\,c_F\bigg[4\,\ln(\bar{\mu}^2/q^2)\bigg] 
+ \frac{g_{RI^{\prime}}^4}{(16\pi^2)^2}\,c_F&\bigg[&\ln^2(\bar{\mu}^2/q^2)\left(-N_f+8\,c_F+\frac{13}{2}\,N_c\right) \nonumber \\ 
&&+ \ln(\bar{\mu}^2/q^2)\left(-\frac{58}{9}\,N_f+\frac{421}{9}\,N_c\right)\bigg]
\label{ScalarRenormGreenFnRI}
\end{eqnarray}
Eq.(\ref{ScalarRenormGreenFnRI}) holds also for the case of the pseudoscalar operator: $G^{RI^{\prime}}_{P}=G^{RI^{\prime}}_{S}$.
\begin{eqnarray}
G^{\overline{MS}}_{S}=1 &+& \frac{g_{\overline{MS}}^2}{16\pi^2}\,c_F\bigg[4\,\ln(\bar{\mu}^2/q^2)+6\bigg] \nonumber \\
&+& \frac{g_{\overline{MS}}^4}{(16\pi^2)^2}\,c_F\bigg[\ln^2(\bar{\mu}^2/q^2)\left(-N_f+8\,c_F+\frac{13}{2}\,N_c\right) \nonumber \\ 
&&\quad\qquad\,+ \ln(\bar{\mu}^2/q^2)\left(-\frac{16}{3}\,N_f+24\,c_F+\frac{130}{3}\,N_c\right) \nonumber \\
&&\quad\qquad\,+\left(-\frac{26}{3}\,N_f+\bigg(22+12\zeta(3)\bigg)\,c_F+\left(\frac{227}{3}-24\zeta(3)\right)\,N_c \right)\bigg]
\label{ScalarRenormGreenFnMS}
\end{eqnarray}

\begin{eqnarray}
G^{\overline{MS}}_{P}=1 &+& \frac{g_{\overline{MS}}^2}{16\pi^2}\,c_F\bigg[4\,\ln(\bar{\mu}^2/q^2)+14\bigg] \nonumber \\
&+& \frac{g_{\overline{MS}}^4}{(16\pi^2)^2}\,c_F\bigg[\ln^2(\bar{\mu}^2/q^2)\left(-N_f+8\,c_F+\frac{13}{2}\,N_c\right) \nonumber \\ 
&&\quad\qquad\,+ \ln(\bar{\mu}^2/q^2)\left(-\frac{16}{3}\,N_f+56\,c_F+\frac{130}{3}\,N_c\right) \nonumber \\
&&\quad\qquad\,+\left(-\frac{82}{9}\,N_f+\bigg(134+12\zeta(3)\bigg)\,c_F+\left(\frac{679}{9}-24\zeta(3)\right)\,N_c \right)\bigg]
\label{PseudoRenormGreenFnMS}
\end{eqnarray}

In Figs. (6a,6b), (7a,7b), (8a,8b) we plot
($Z_{\psi}^{L,\overline{MS}}$, $Z_{\psi}^{L,RI^{\prime}}$), 
($Z_S^{L,\overline{MS}}$, $Z_S^{L,RI^{\prime}}$) and 
($Z_P^{L,\overline{MS}}$, $Z_P^{L,RI^{\prime}}$), respectively, 
as a function of $c_{\rm SW}$. In practice, of course, only specific values
of $c_{\rm SW}$ are relevant, in the range $1\le c_{\rm SW} \le 1.8$,
corresponding to perturbative or non-perturbative determinations.
For definiteness, we have set 
$N_c=3$, $\bar{\mu}=1/a_{_{\rm L}}$ and
$\beta_{\rm o}\equiv 2N_c/g_{\rm o}^2=6.0$. Our results up to two loops 
for each $Z$ are shown for both $N_f=0$ and $N_f=2$, and 
compared to the corresponding one-loop results. Furthermore, in  
the scalar case, we also present the two-loop result for the flavor
singlet operator.

In Fig.9 we present, on the same plot, the values of 
$Z_{\psi}^{L,\overline{MS}}$, $Z_S^{L,\overline{MS}}$, 
$Z_P^{L,\overline{MS}}$ and $Z_{S,\,singlet}^{L,\overline{MS}}$ 
up to 2 loops, versus $c_{\rm SW}$. 
We have chosen $N_c=3$, $\bar{\mu}=1/a_{_{\rm L}}$, $N_f=2$ 
and $\beta_{\rm o}=5.3$. The corresponding results in the 
$RI^{\prime}$ scheme are plotted in Fig.10. 

There are a number of non-perturbative (NP) estimates of
renormalization constants in the literature, in the $RI^{\prime}$
scheme (see, e.g.,~\cite{Crisafulli,Becirevic2,Becirevic3}) and in the
Schr\"odinger functional scheme~\cite{DellaMorte2}.
Our 2-loop results still differ from NP results in $RI^{\prime}$, and
this leaves open the possibility that higher loop effects may still be
important, even though the perturbative series shows reasonable signs
of convergence. 
A putative reason for this difference is the fact that the bare coupling
constant $g_\circ$ is known not to be a good expansion parameter.
One may also
express the renormalization functions in terms of the renormalized
couplings: $g_{\overline{MS}}$ or $g_{RI^{\prime}}$. The resulting
expressions for $Z_S^{L,Y}$, $Z_P^{L,Y}$ and $Z_{S,\,singlet}^{L,Y}$
($Y=\overline{MS}\,,\,RI^{\prime}$) as a function of $c_{\rm SW}$, are
shown in Figs.11 and 12, for the same values for $N_c$, $N_f$,
$\bar{\mu}$ and $\beta_\circ$ as in Figs.9 and 10. For values of the
clover parameter beyond its typical range, $c_{\rm SW} \ge 1.8$, the
behavior of the renormalization functions shows signs of instability
at the scale $\bar{\mu}=1/a_{_{\rm L}}$\,.
There exist also
several alternative definitions of an effective coupling in the
literature; one should be aware, however, that the use of many of
these definitions (coming, e.g., from boosted perturbation theory)
can only be justified for 1-loop quantities, not beyond. For this
reason, we have preferred to provide the bare results in this paper,
leaving to the reader the straightforward task of converting
these results to their favorite scheme. 

\newpage

\begin{center}
\psfig{figure=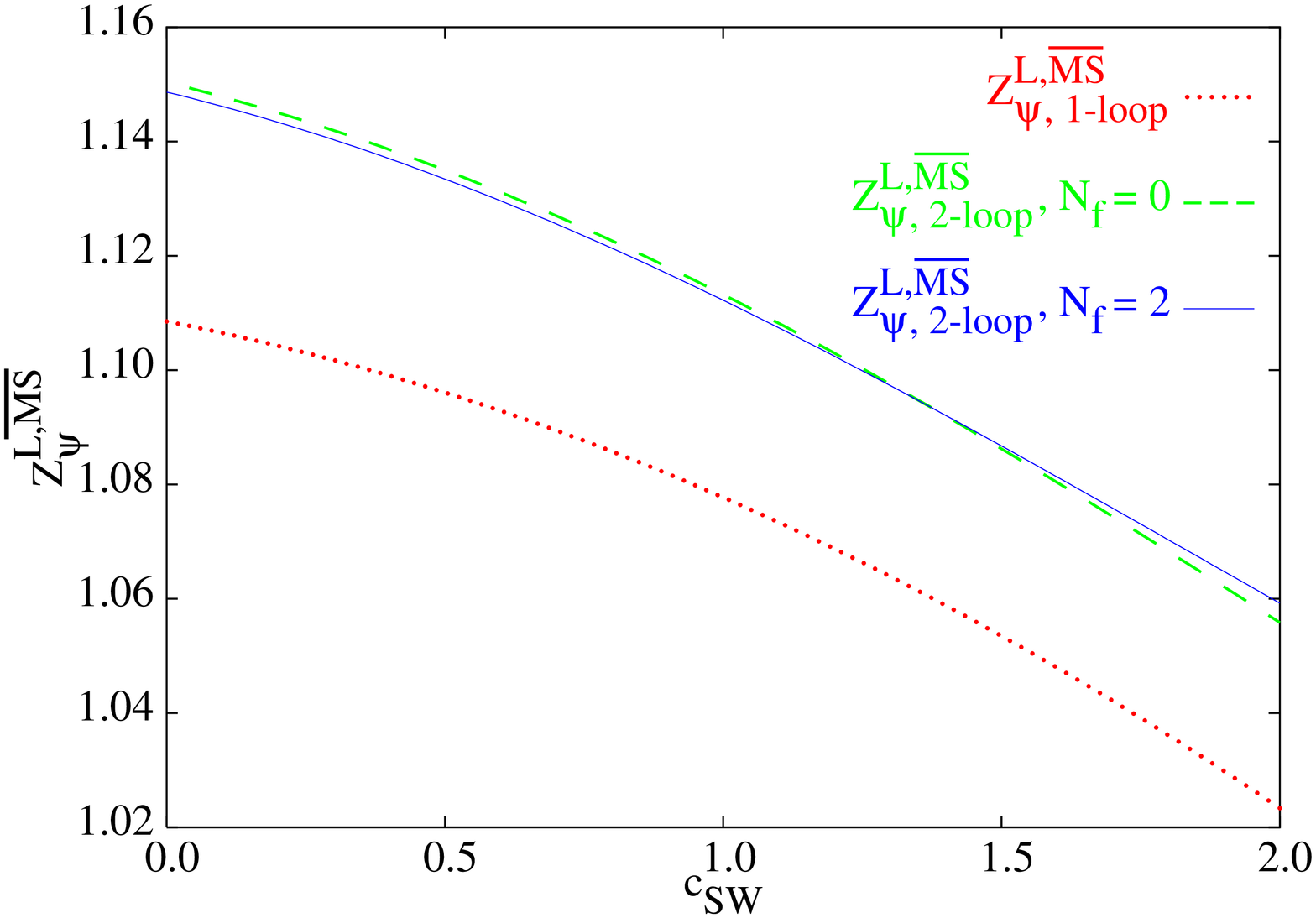,scale=0.50, angle=-90}
\vskip 4mm
{\small FIG. 6a. $Z_{\psi}^{L,\overline{MS}}(a_{_{\rm L}}\bar{\mu})$  versus $c_{{\rm SW}}$ 
($N_c=3$, $\bar{\mu}=1/a_{_{\rm L}}$, $\beta_{\rm o}=6.0$). Results up to 2 loops are shown 
for $N_f=0$ (dashed line) and $N_f=2$ (solid line); one-loop results are plotted 
with a dotted line.}
\end{center}

\begin{center}
\psfig{figure=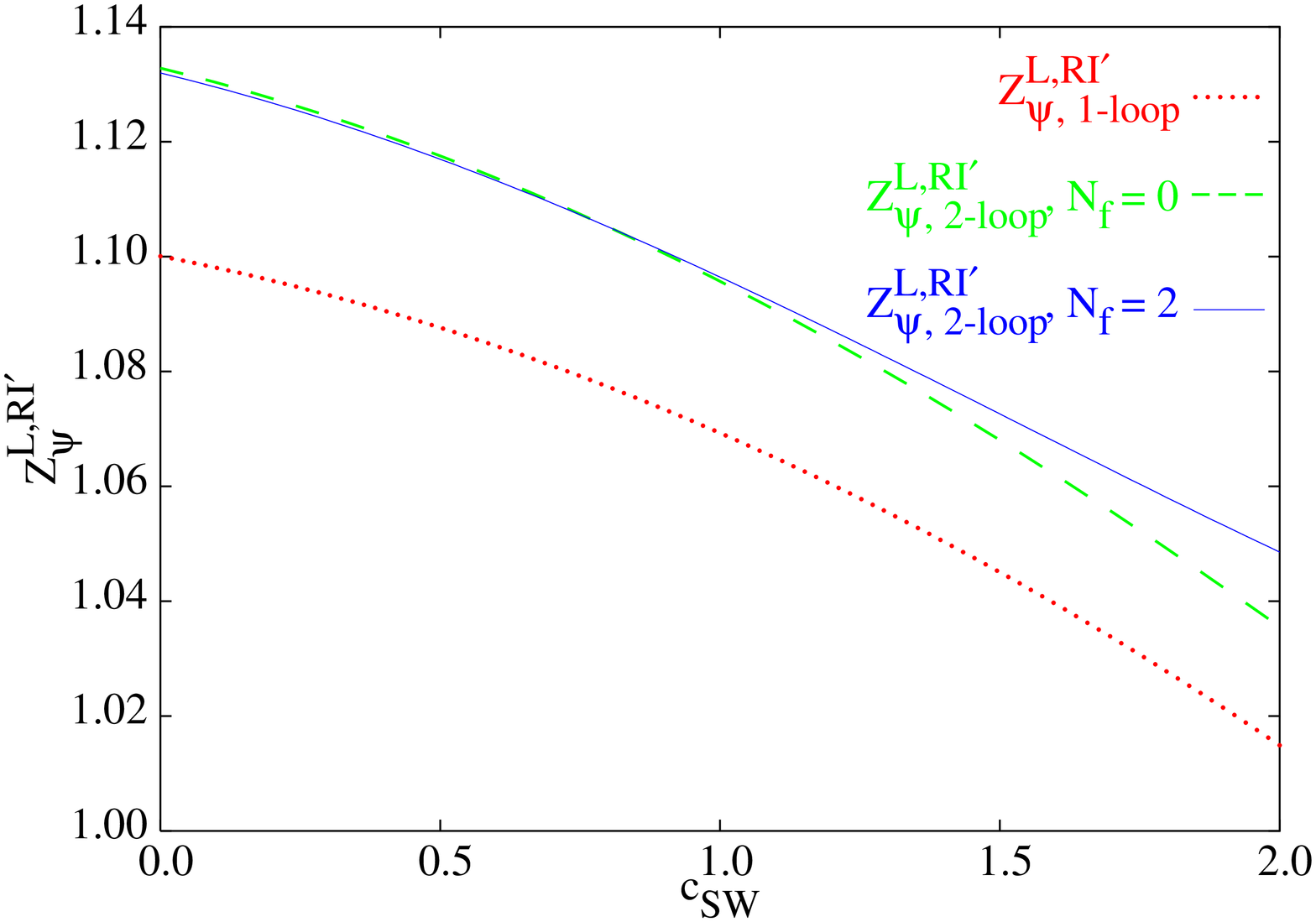,scale=0.50,angle=-90}
\vskip 4mm
{\small FIG. 6b. $Z_{\psi}^{L,RI^{\prime}}(a_{_{\rm L}}\bar{\mu})$ versus $c_{{\rm SW}}$ 
($N_c=3$, $\bar{\mu}=1/a_{_{\rm L}}$, $\beta_{\rm o}=6.0$). Same notation as in FIG.6a.}
\end{center}

\newpage

\begin{center}
\psfig{figure=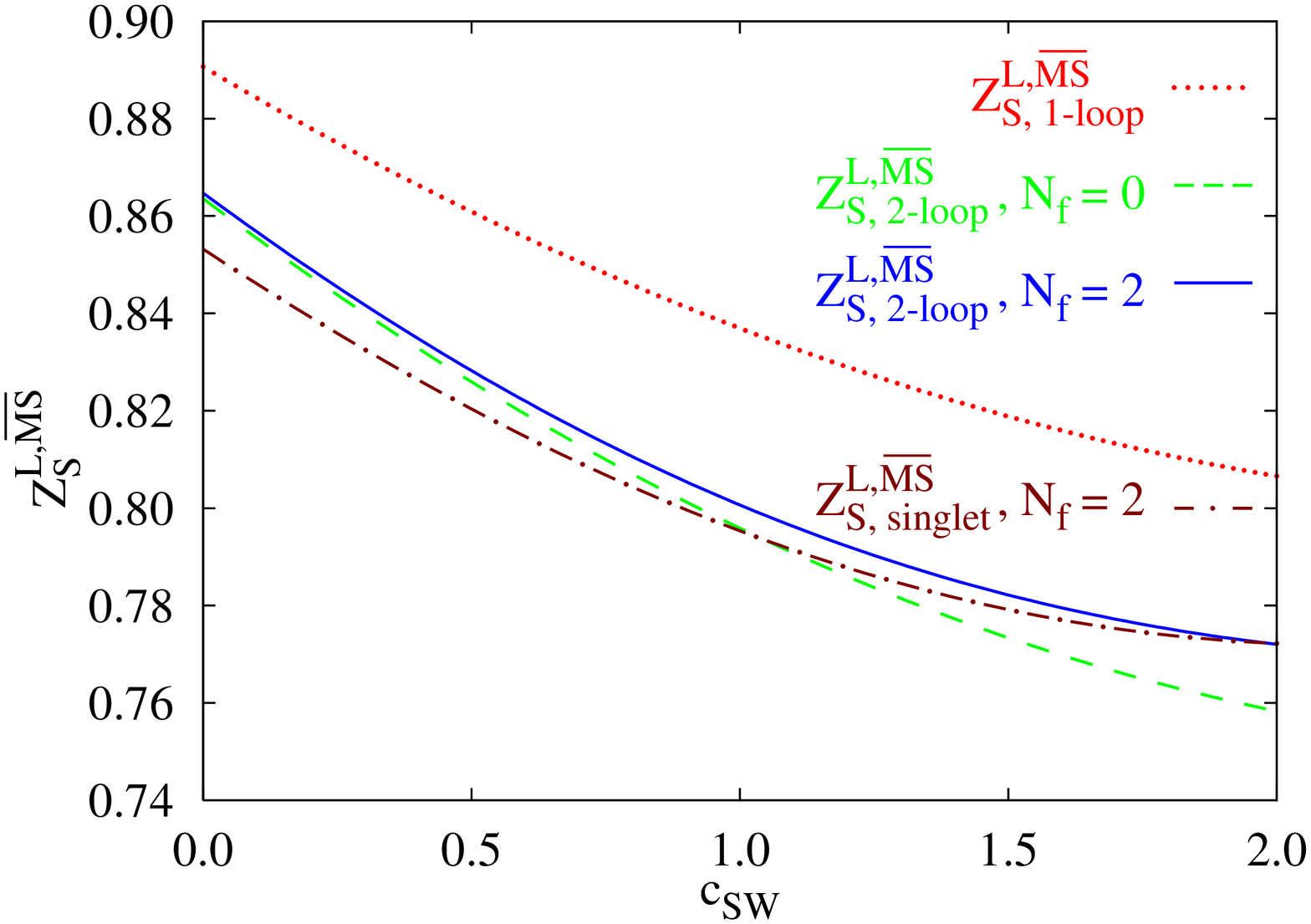,scale=0.50, angle=-90}
\vskip 4mm
{\small FIG. 7a. $Z_S^{L,\overline{MS}}(a_{_{\rm L}}\bar{\mu})$ 
versus $c_{{\rm SW}}$ 
($N_c=3$, $\bar{\mu}=1/a_{_{\rm L}}$, $\beta_{\rm o}=6.0$). 
Results up to 2 loops, for the flavor non-singlet operator, 
are shown for $N_f=0$ (dashed line) and $N_f=2$ (solid line); 
2-loop results for the flavor singlet operator, for $N_f=2$, 
are plotted with a dash-dotted line; one-loop results are 
plotted with a dotted line.}
\end{center}

\begin{center}
\psfig{figure=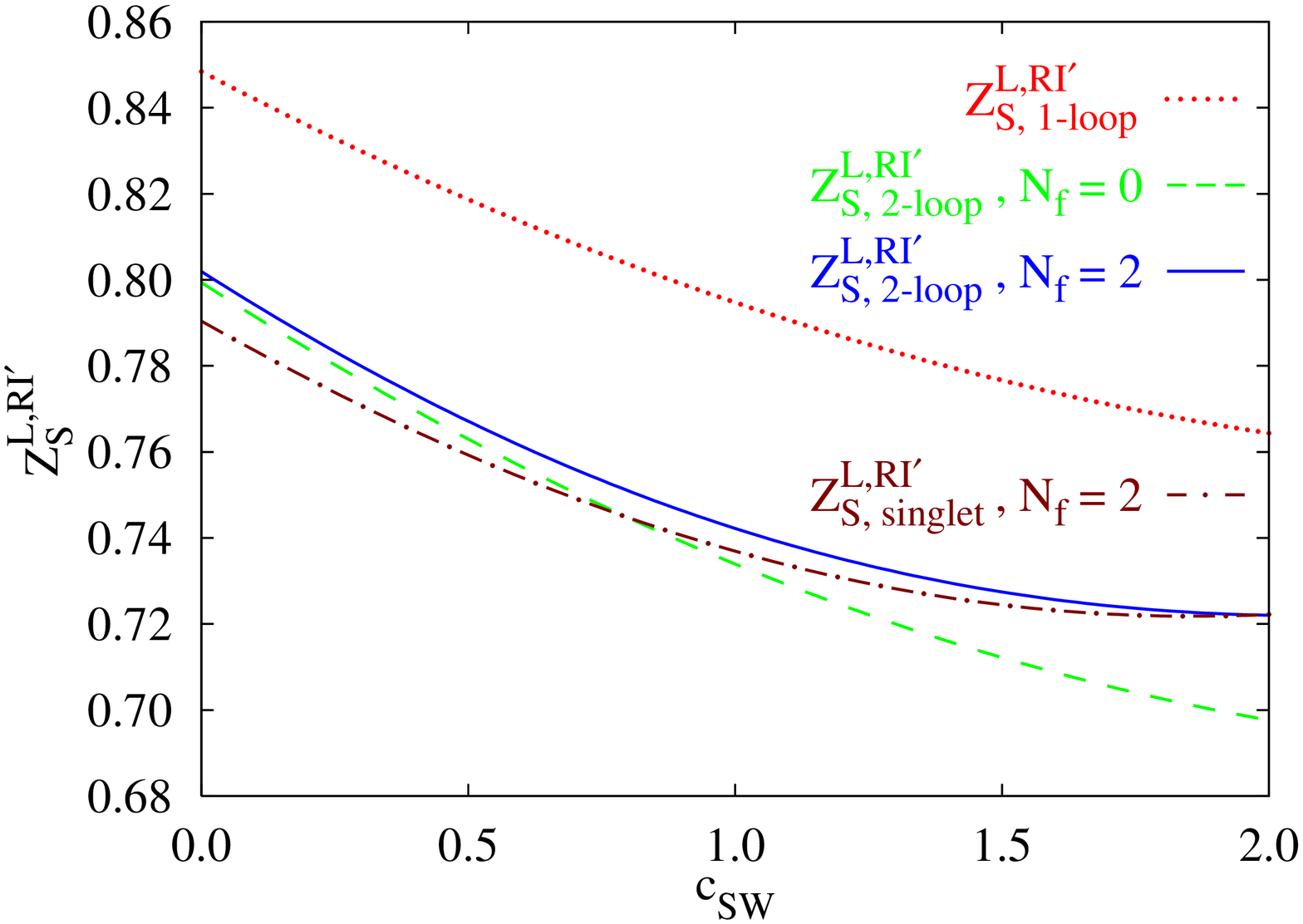,scale=0.50,angle=-90}
\vskip 4mm
{\small FIG. 7b. $Z_S^{L,RI^{\prime}}(a_{_{\rm L}}\bar{\mu})$ 
versus $c_{{\rm SW}}$ 
($N_c=3$, $\bar{\mu}=1/a_{_{\rm L}}$, $\beta_{\rm o}=6.0$). 
Same notation as in FIG.7a.}
\end{center}

\newpage

\begin{center}
\psfig{figure=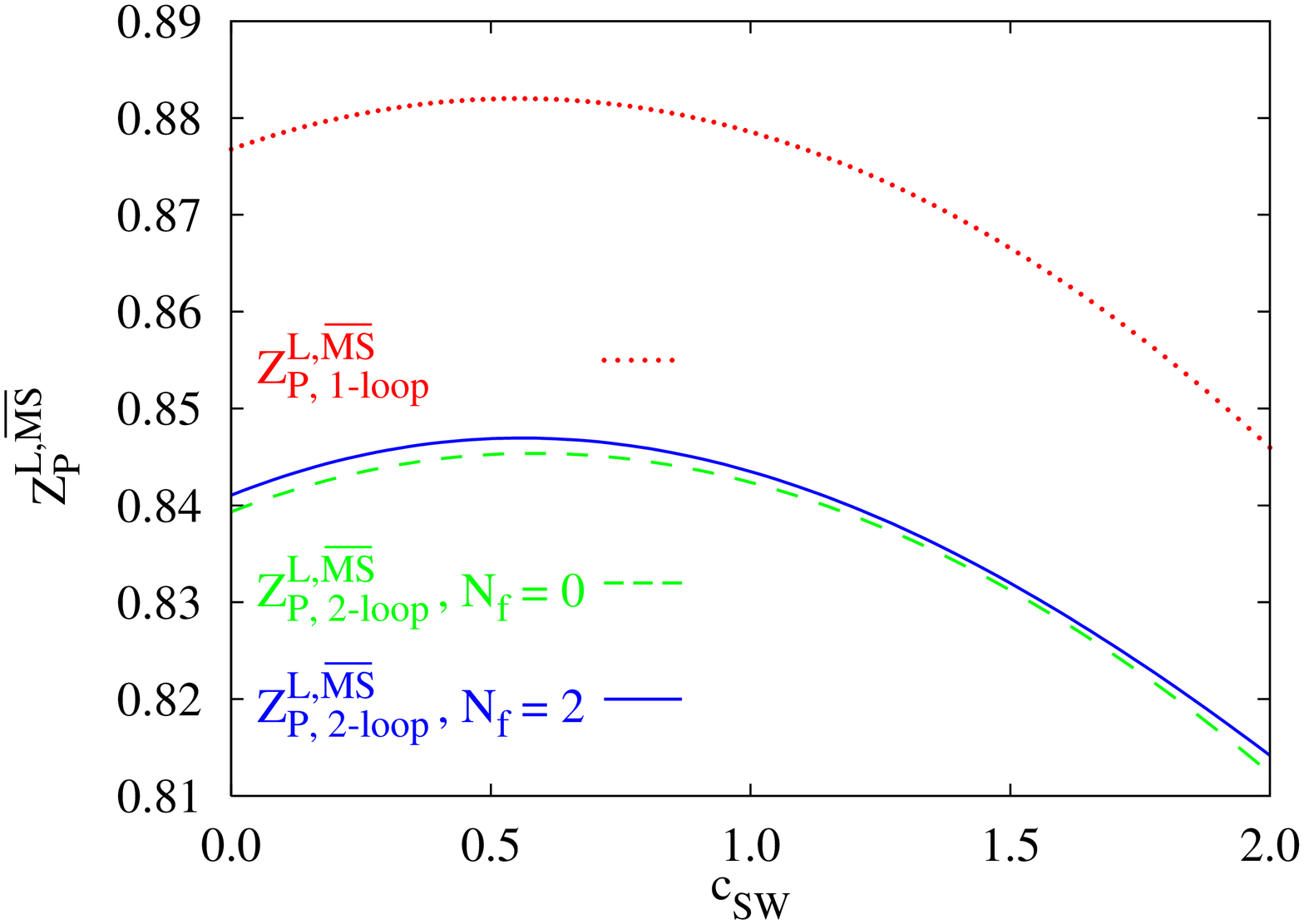,scale=0.50, angle=-90}
\vskip 4mm
{\small FIG. 8a. $Z_P^{L,\overline{MS}}(a_{_{\rm L}}\bar{\mu})$ 
versus $c_{{\rm SW}}$ 
($N_c=3$, $\bar{\mu}=1/a_{_{\rm L}}$, $\beta_{\rm o}=6.0$). 
Results up to 2 loops are shown for $N_f=0$ (dashed line) and 
$N_f=2$ (solid line); one-loop results are plotted with a dotted line.}
\end{center}

\begin{center}
\psfig{figure=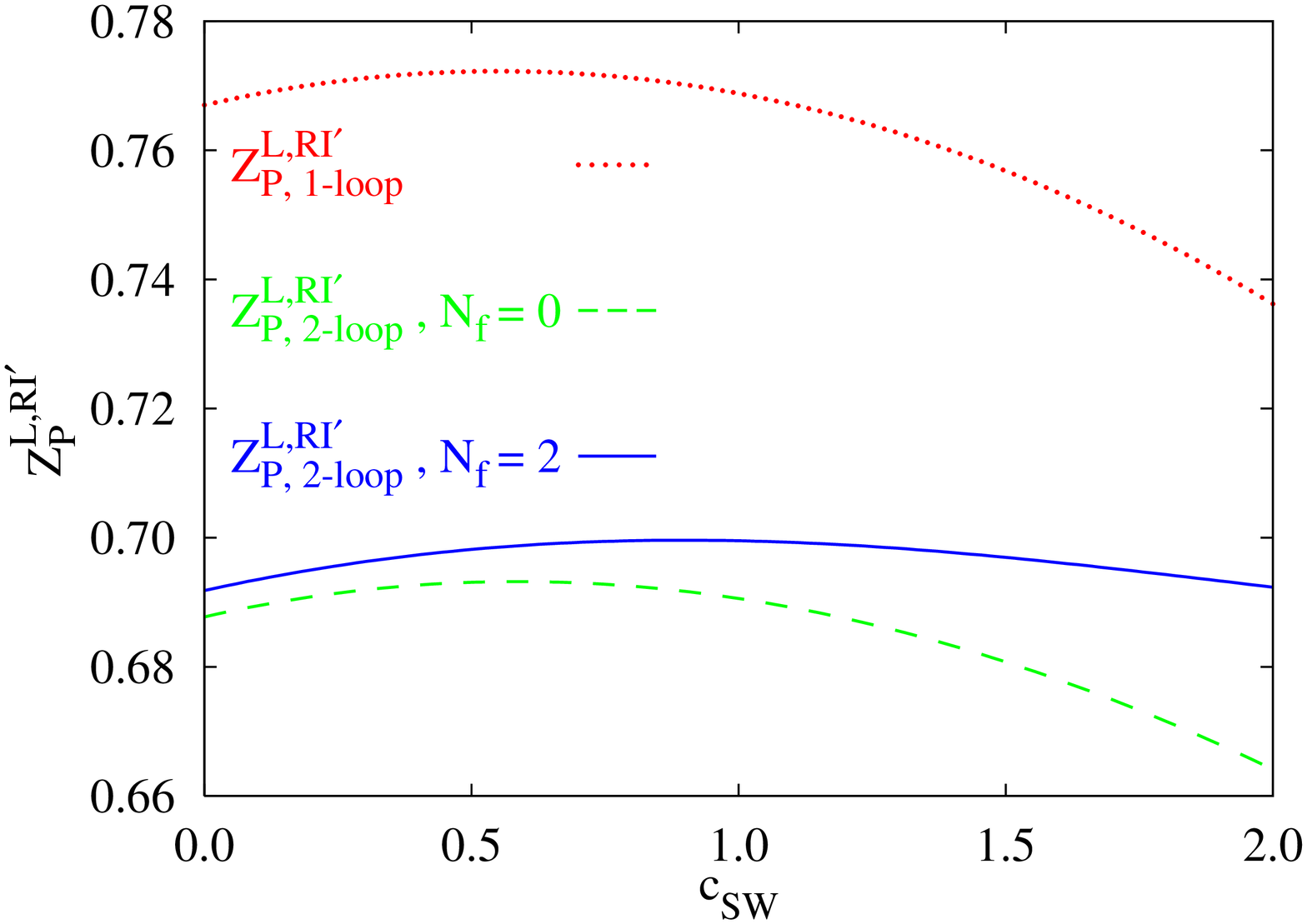,scale=0.50,angle=-90}
\vskip 4mm
{\small FIG. 8b. $Z_P^{L,RI^{\prime}}(a_{_{\rm L}}\bar{\mu})$ 
versus $c_{{\rm SW}}$ 
($N_c=3$, $\bar{\mu}=1/a_{_{\rm L}}$, $\beta_{\rm o}=6.0$). 
Same notation as in FIG.8a.}
\end{center}

\newpage

\begin{center}
\psfig{figure=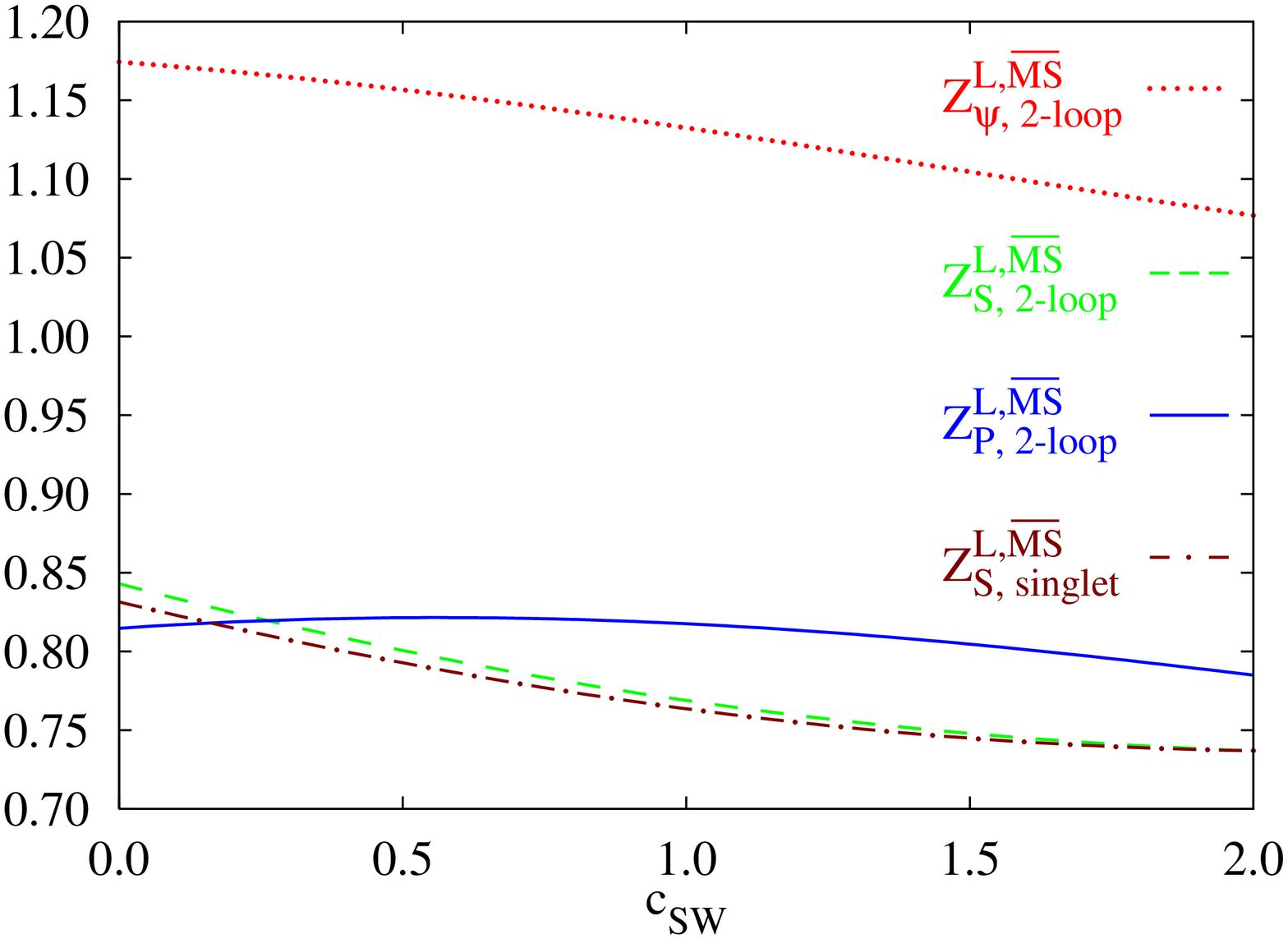,scale=0.50, angle=-90}
\vskip 4mm
{\small FIG. 9. \Red{$Z_{\psi}^{L,\overline{MS}}$} (dotted line), 
\Green{$Z_S^{L,\overline{MS}}$} (dashed line), 
\Blue{$Z_P^{L,\overline{MS}}$} (solid line) and 
\Brown{$Z_{S,\,singlet}^{L,\overline{MS}}$} (dash-dotted line)
up to 2 loops, versus $c_{\rm SW}$ ($N_c=3$, $\bar{\mu}=1/a_{_{\rm L}}$, 
$N_f=2$, $\beta_{\rm o}=5.3$).}
\end{center}

\begin{center}
\psfig{figure=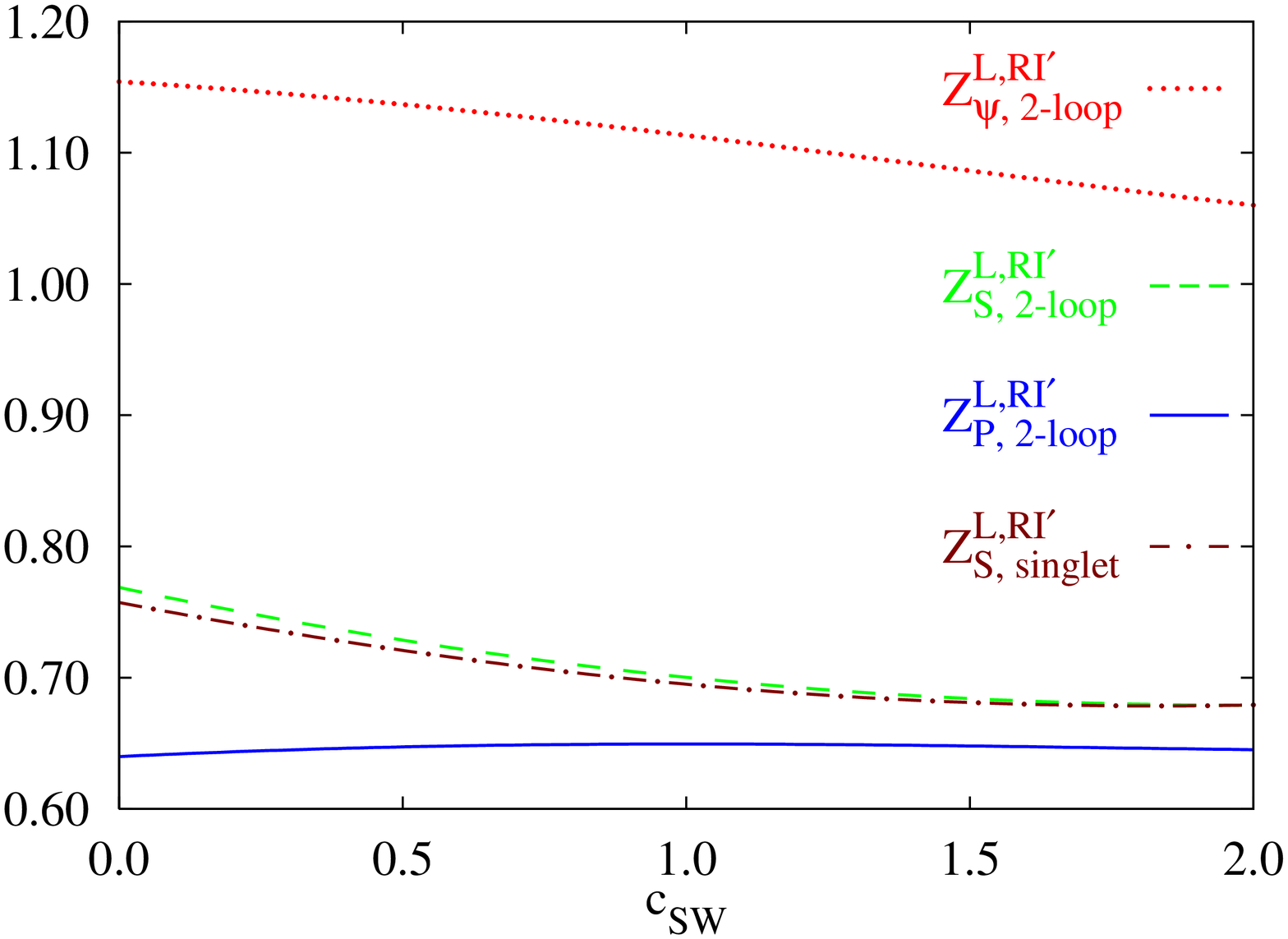,scale=0.50,angle=-90}
\vskip 4mm
{\small FIG. 10. \Red{$Z_{\psi}^{L,RI^{\prime}}$} (dotted line), 
\Green{$Z_S^{L,RI^{\prime}}$} (dashed line), 
\Blue{$Z_P^{L,RI^{\prime}}$} (solid line) and
\Brown{$Z_{S,\,singlet}^{L,RI^{\prime}}$} (dash-dotted line)
up to 2 loops, versus $c_{\rm SW}$ ($N_c=3$, $\bar{\mu}=1/a_{_{\rm L}}$, 
$N_f=2$, $\beta_{\rm o}=5.3$).}
\end{center}

\begin{center}
\psfig{figure=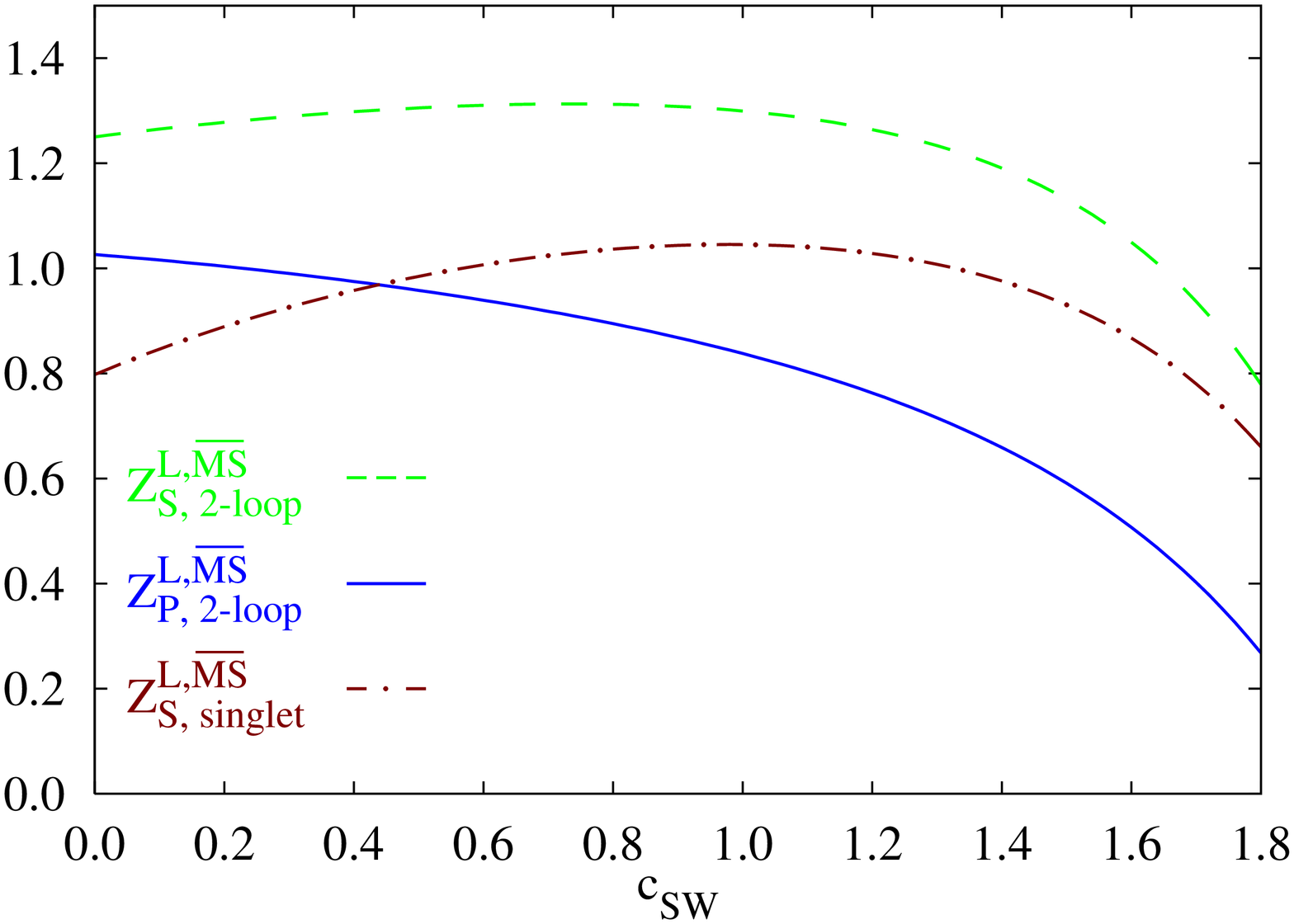,scale=0.52, angle=-90}
\vskip 4mm
{\small FIG. 11. 2-loop renormalization functions
  \Green{$Z_S^{L,\overline{MS}}$} (dashed line),  
\Blue{$Z_P^{L,\overline{MS}}$} (solid line) and 
\Brown{$Z_{S,\,singlet}^{L,\overline{MS}}$} (dash-dotted line)
expressed in terms of the renormalized coupling constant
$g_{\overline{MS}}$, versus 
$c_{\rm SW}$ ($N_c=3$, $\bar{\mu}=1/a_{_{\rm L}}$,  
$N_f=2$, $\beta_{\rm o}=5.3$).}
\end{center}
\begin{center}
\vskip -0.55cm 
\psfig{figure=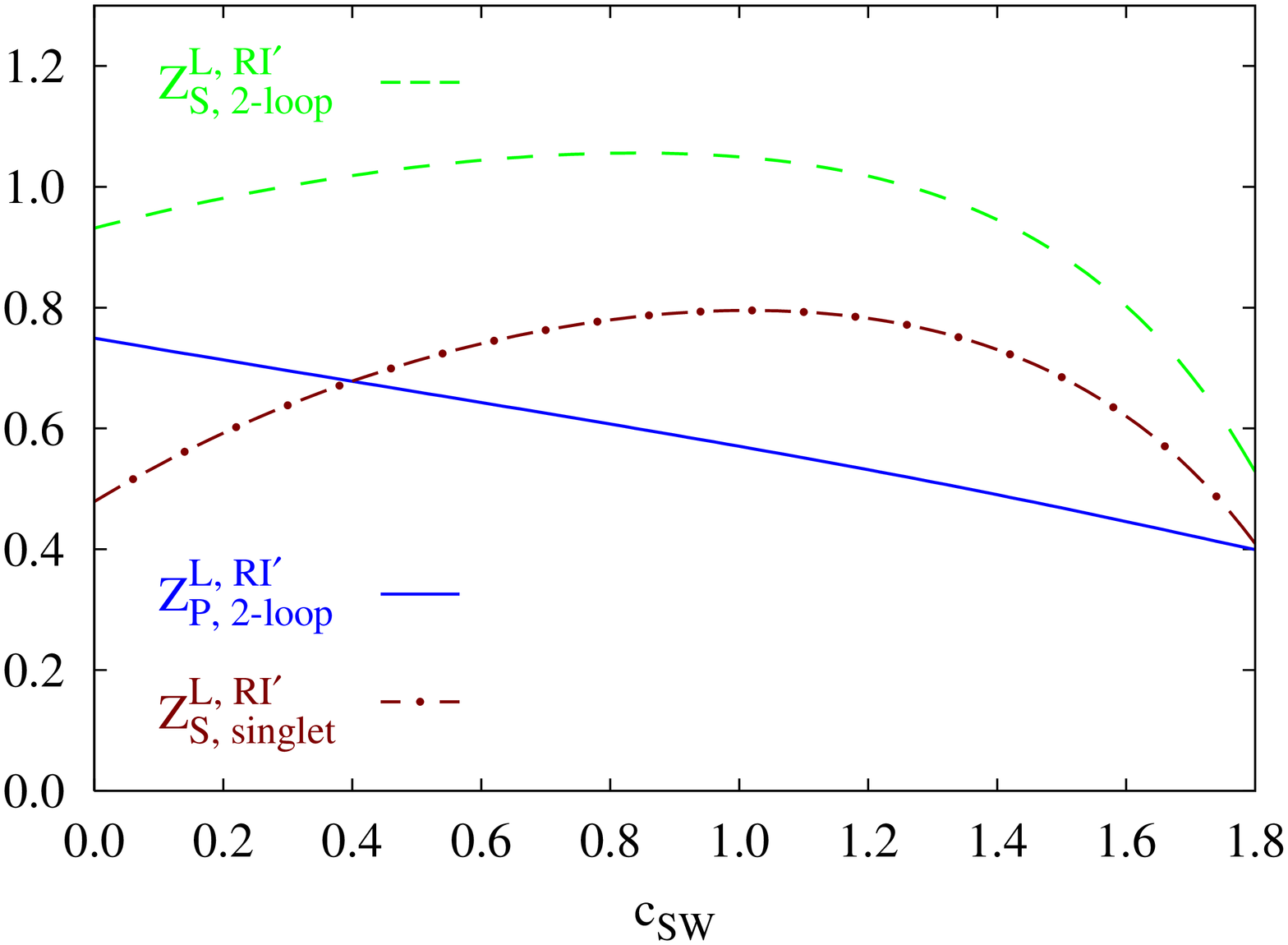,scale=0.52,angle=-90}
\vskip 4mm
{\small FIG. 12. 2-loop renormalization functions
\Green{$Z_S^{L,RI^{\prime}}$} (dashed line), 
\Blue{$Z_P^{L,RI^{\prime}}$} (solid line) and
\Brown{$Z_{S,\,singlet}^{L,RI^{\prime}}$} (dash-dotted line)
expressed in terms of the renormalized coupling constant
$g_{RI^{\prime}}$, versus  
$c_{\rm SW}$ ($N_c=3$, $\bar{\mu}=1/a_{_{\rm L}}$, 
$N_f=2$, $\beta_{\rm o}=5.3$).}
\end{center}


\section{Discussion}
\label{Discussion}

As can be seen from Figs.6a-8b, all 2-loop
renormalization functions differ from 1-loop values in a significant
way; this difference should be taken into account in MC simulations, in order
to reduce systematic error.  At the same time, 2-loop contributions are consistently smaller
than 1-loop contributions, indicating that the (asymptotic) perturbative series are
under control. 

The dependence on the clover parameter
$c_{\rm SW}$ is also quite pronounced. In the present work, $c_{\rm SW}$ was left as
a free parameter; its optimal value, as dictated by
${\cal{O}}(a_{_{\rm L}})$ improvement, has been estimated both
non-perturbatively~\cite{Luscher1996} and perturbatively (to
1-loop)~\cite{SW}.

Our results regard both the flavor nonsinglet and singlet operators.
For the pseudoscalar operator, these cases coincide, just as in
dimensional regularization. The scalar operator, on the other hand,
receives an additional finite ($a_{_{\rm L}}\bar{\mu}$ independent)
contribution in the flavor singlet case.
$Z_{S,\,singlet}$ is seen to be equal to the fermion mass
renormalization $Z_m$\,, which is an essential ingredient in the
computation of quark masses.

We note also that, in dimensional regularization, both the scalar
and pseudoscalar flavor singlet operators renormalize in the same way
as their non-singlet counterparts, for mass independent
renormalization schemes. Consequently, the conversion factors $C_S$ and $C_P$, as
well as $Z_5$, stay the same for flavor singlets.

A breakdown of our results on a {\em per
  diagram} basis has not been presented here, due to lack of space; it
  is available from the authors upon request.

The 2-loop computation of the
renormalization functions for the Vector, Axial and Tensor bilinears
is work currently in progress.

Besides the strictly local definitions of
fermion bilinears, $\bar{\psi}\Gamma\psi$, one can consider a family
of more extended operators (see, e.g., \cite{Luscher1996}), with the same classical continuum limit,
as dictated by ${\cal O}(a_{_{\rm L}})$ improvement. The
renormalization of these extended operators involves more Feynman
diagrams, since their vertices may also contain gluon lines; however,
the computation is actually less cumbersome, since all additional
contributions are now free of superficial divergences. We will be
reporting the results of this computation in a future work.

\newpage
\begin{center}
{\bf APPENDIX: Fermions in an arbitrary representation}
\end{center}

Our results for $Z_\psi$, $Z_S$, $Z_P$, 
Eqs.(\ref{Zpsi2loopRI}, \ref{ZS2loopRI}, \ref{ZP2loopRI}), can be
easily generalized to an action with Wilson/clover fermions in an
arbitrary representation $R$, of dimensionality $d_R$\,.

In this case, the gluon part of the action remains the same, while all
link variables appearing in the fermion part of the action assume the
form:
\begin{equation}
U_{x,\,x+\mu} = {\rm exp}(i\,g_0\,A^a_\mu(x)\,T^a)\quad\longrightarrow\quad
U_{x,\,x+\mu} = {\rm exp}(i\,g_0\,A^a_\mu(x)\,T^a_R)
\end{equation}
Using standard notation and conventions, the generators $T^a$ in
the fundamental representation satisfy:
\begin{equation}
[T^a,T^b] = i\,f^{abc}\,T^c,\quad \sum_aT^aT^a 
\equiv \openone\,c_F = \openone\,{N_c^2-1\over 2N_c},
\quad {\rm tr}(T^aT^b)\equiv\delta^{ab}\,t_F=
\delta^{ab}\,{1\over2}
\end{equation}
In the representation $R$ we have:
\begin{equation}
[T^a_R,T^b_R] = i\,f^{abc}\,T^c_R,\quad \sum_aT^a_RT^a_R 
\equiv \openone\,c_R,\quad
{\rm tr}(T^a_RT^b_R)\equiv\delta^{ab}\,t_R
\end{equation}
where: $t_R=(d_R\,c_R)/(N_c^2-1)$.

For the 1-loop quantities, Eqs.(\ref{ZA1loop}, \ref{Zg1loop}),
converting to the representation $R$ is a straightforward
substitution:
\begin{equation}
N_f \longrightarrow N_f \cdot (2\,t_R)
\label{tR}
\end{equation}
and, in addition, for Eqs.(\ref{Zpsi1loopRI}-\ref{ZP1loopMS}):
\begin{equation}
c_F \longrightarrow c_R
\label{cR}
\end{equation}
Aside from these changes, all algebraic expressions (and the numerical
coefficients resulting from loop integrations) remain the same.

A similar reasoning applies to the 2-loop quantities in
Eqs.(\ref{Zpsi2loopRI}, \ref{ZS2loopRI}, \ref{ZP2loopRI}): For most
diagrams, once their value is expressed as a linear
combination of $c_F^2$, $c_F N_c$ and $c_F N_f$\,, it suffices to
apply substitutions (\ref{tR}) and (\ref{cR}). The only exceptions are
diagrams containing a gluon tadpole [diagrams 
7, 14 (Fig.2); diagram 3 (Fig.4); 1-loop diagrams, when expressed in
terms of $a_{RI^{\prime}}$, $\alpha_{RI^{\prime}}$ by means of $Z_g$,
$Z_A$]: In these cases, only one power of $c_F$ should be changed to
$c_R$\,; a possible additional power of $c_F$ originates from a gluon
tadpole and should stay as is. This peculiarity implies that, in order
to perform the substitutions as described above, one must start from the
{\em per diagram} breakdown of 2-loop results.
To avoid presenting a lengthy breakdown, we apply, instead,
substitutions (\ref{tR}) and (\ref{cR}) indiscriminately on  
Eqs.(\ref{Zpsi2loopRI}, \ref{ZS2loopRI}, \ref{ZP2loopRI});
consequently, we must
then add a correction term, as follows:

\newpage
\begin{eqnarray}
Z_{\psi}^{L,RI^{\prime}}\big|_R &=&
Z_{\psi}^{L,RI^{\prime}}\big|_{c_F\to c_R\,,\ N_f\to 2 N_f\,t_R}
\nonumber\\
&&+\frac{g_\circ^4}{(16\pi^2)^2}\,c_R\,(c_R-c_F) \cdot \big[ -4 \pi^2 \ln(a_{_{\rm L}}^2 \bar{\mu}^2)
  -467.9141661(2) \nonumber\\
&&\hskip 3.5cm + 88.7817709(1)\,c_{{\rm SW}} +
  55.1618942(2)\,c_{{\rm SW}}^2 \big]\\
Z_S^{L,RI^{\prime}}\big|_R &=&
Z_S^{L,RI^{\prime}}\big|_{c_F\to c_R\,,\ N_f\to 2 N_f\,t_R}
\nonumber\\
&&+\frac{g_\circ^4}{(16\pi^2)^2}\,c_R\,(c_R-c_F) \cdot \big[ -12 \pi^2 \ln(a_{_{\rm L}}^2 \bar{\mu}^2)
  +708.732752(6) \nonumber\\
&&\hskip 3.5cm + 305.48068(1)\,c_{{\rm SW}} 
  -54.495244(2)\,c_{{\rm SW}}^2 \big]\\
Z_P^{L,RI^{\prime}}\big|_R &=&
Z_P^{L,RI^{\prime}}\big|_{c_F\to c_R\,,\ N_f\to 2 N_f\,t_R}
\nonumber\\
&&+\frac{g_\circ^4}{(16\pi^2)^2}\,c_R\,(c_R-c_F) \cdot \big[ -12 \pi^2 \ln(a_{_{\rm L}}^2 \bar{\mu}^2)
  +1089.424358(4)\nonumber\\
&&\hskip 3.5cm -88.7817709(1)\,c_{{\rm SW}} 
  +80.378675(2)\,c_{{\rm SW}}^2 \big]
\end{eqnarray}
[Actually, the reader could arrive at these results without knowledge
  of the {\em per diagram} breakdown, by virtue of the following fact:
  All `exceptional' powers of $c_F$ cancel out of
  $Z_{\psi}^{L,RI^{\prime}}$, $Z_S^{L,RI^{\prime}}$, $Z_P^{L,RI^{\prime}}$, 
if these are
expressed in terms of the renormalized coupling constant
$a_{RI^{\prime}}$\,. Thus, one may: 
\begin{itemize}
\item[$\bullet$] Express 
Eqs.(\ref{Zpsi2loopRI}, \ref{ZS2loopRI}, \ref{ZP2loopRI})
in terms of $g_{RI^{\prime}}$\, by means of $g_{\rm o} =
(Z_g^{L,RI^{\prime}})\,g_{RI^{\prime}}$\,, with $Z_g^{L,RI^{\prime}}$
in the fundamental representation (Eq.(\ref{Zg1loop})) 
\item[$\bullet$] Apply substitutions (\ref{tR}), (\ref{cR}) throughout
\item[$\bullet$] If desired, reexpress everything in terms of $g_{\rm o}$ 
(using $(Z_g^{L,RI^{\prime}})^{-1}$ from Eq.(\ref{Zg1loop}), with
$N_f\to 2N_ft_R$ and $c_F$ as is)
\end{itemize}
No correction terms are necessary in this procedure.]

\bigskip\noindent
{\bf Acknowledgements: } We would like to thank J. A. Gracey and S. A.
Larin for private communication regarding their continuum results. 
This work is supported in part by the
Research Promotion Foundation of Cyprus 
(Proposal Nr: $\rm ENI\Sigma X$/0506/17).


\end{document}